\documentclass{aastex63}

\newcommand{\jw}{\color{black}}
\newcommand{\chen}{\color{black}}

\newcommand{\cdc}{\color{black}}
\newcommand{\CDC}{\color{black}}

\usepackage{amsmath}
\usepackage{amssymb}
\usepackage{color}
\usepackage{url}
\usepackage{ulem}
\usepackage{multirow}
\usepackage{graphics}
\usepackage{longtable}
\usepackage{graphicx}
\usepackage{bm}
\usepackage{appendix}

\begin{document}

\title{{\jw Planets Across Space and Time (PAST). \uppercase\expandafter{\romannumeral3}. \\ Morphology of the Planetary Radius Valley as a Function of Stellar Age and Metallicity in the Galactic Context Revealed by the LAMOST-Gaia-Kepler Sample}

}

\correspondingauthor{Ji-Wei Xie}
\email{jwxie@nju.edu.cn}


\author{Di-Chang Chen$^{*}$}
\affiliation{School of Astronomy and Space Science, Nanjing University, Nanjing 210023, China}
\affiliation{Key Laboratory of Modern Astronomy and Astrophysics, Ministry of Education, Nanjing 210023, China}
\renewcommand{\thefootnote}{}
\footnotetext{* LAMOST Fellow}

\author{Ji~Wei Xie}
\affiliation{School of Astronomy and Space Science, Nanjing University, Nanjing 210023, China}
\affiliation{Key Laboratory of Modern Astronomy and Astrophysics, Ministry of Education, Nanjing 210023, China}

\author{Ji-Lin. Zhou}
\affiliation{School of Astronomy and Space Science, Nanjing University, Nanjing 210023, China}
\affiliation{Key Laboratory of Modern Astronomy and Astrophysics, Ministry of Education, Nanjing 210023, China}

\author{Jia-Yi Yang}
\affiliation{School of Astronomy and Space Science, Nanjing University, Nanjing 210023, China}
\affiliation{Key Laboratory of Modern Astronomy and Astrophysics, Ministry of Education, Nanjing 210023, China}


\author{Subo Dong}
\affiliation{Kavli Institute for Astronomy and Astrophysics, Peking University, Beijing 100871, China}


\author{Zi Zhu}
\affiliation{School of Astronomy and Space Science, Nanjing University, Nanjing 210023, China}
\affiliation{Key Laboratory of Modern Astronomy and Astrophysics, Ministry of Education, Nanjing 210023, China}

\author{Zheng Zheng}
\affiliation{Department of Physics and Astronomy, University of Utah, Salt Lake City, UT 84112}

\author{Chao Liu}
\affiliation{National Astronomical Observatories, Chinese Academy of Sciences, Beijing 100012, China}

\author{Weikai Zong}
\affiliation{Department of Astronomy, Beijing Normal University, Beijing 100875, China}

\author{Ali Luo}
\affiliation{National Astronomical Observatories, Chinese Academy of Sciences, Beijing 100012, China}




\begin{abstract}
The radius valley, a dip in the radius distribution of exoplanets at $\sim 1.9 R_{\oplus}$ separates compact rocky Super-Earths and Sub-Neptunes with lower density.
Various hypotheses have been put forward to explain the radius valley.
Characterizing the radius valley morphology and its correlation to stellar properties will provide crucial observation constraints on {\jw its origin mechanism} and deepen the understanding of planet formation and evolution. 
In this paper, the third part of the Planets Across the Space and Time (PAST) series, using the LAMOST-Gaia-Kepler catalog, we perform a {\jw systematical investigation into {\CDC how the radius valley morphology varies} in the Galactic context, i.e., thin/thick galactic disks, stellar age and metallicity abundance ($\rm [Fe/H]$ and $\rm [\alpha/Fe]$).}
{\jw We find that (1) The valley becomes more prominent with the increase of both age and $\rm [Fe/H]$. 
(2) The number ratio of super-Earths to sub-Neptunes monotonically increases with age but decreases with $\rm [Fe/H]$ and $\rm [\alpha/Fe]$.
(3) The average radius of planets above the valley ($2.1-6 R_\oplus$) decreases with age but increases with $\rm [Fe/H]$.
(4) In contrast, the average radius of planets below the valley ($R<1.7 R_\oplus$) is broadly independent on age and metallicity.
Our results demonstrate that the valley morphology as well as the whole planetary radius distribution evolves on a long timescale of giga-years, and metallicities (not only Fe but also other metal elements, e.g., Mg, Si, Ca, Ti) play important roles in planet formation and in the long term planetary evolution.

}

\end{abstract}

\section{introduction}

During the past quarter century, over 4,000 planets have been identified and thousands of candidates to be confirmed \citep[NASA Exoplanet Archive, EA hereafter;][] {2013PASP..125..989A}. 
Large sample of known exoplanets have opened doors to exoplanet statistical studies.
{\jw One of crucial questions is how planet properties depend on their host star properties, which provides important insights on understanding planet formation and evolution \citep{2018haex.bookE.153M,2021arXiv210302127Z}.}
{\jw 
For example, it has been well established that the occurrence rate of giant planets is  strongly correlated to the stellar metallicity \citep{2005ApJ...622.1102F,2015AJ....149...14W}, providing key evidence to the core-accretion model on the giant planet formation \citep[e.g.,][]{1993ARA&A..31..129L,1996Icar..124...62P,2004ApJ...604..388I,2004ApJ...616..567I}.
Furthermore, smaller planets of short orbital periods (e.g., hot Neptunes and super-Earths) are also found to be preferentially around metal-rich stars \citep{2015AJ....149...14W,2018PNAS..115..266D,2018AJ....155...89P}, suggesting metallicity plays an important role in the planet formation and orbital migration. 
}


{\jw Recently, with the large planet sample provided by Kepler, statistical studies have revealed another important planetary feature, i.e., a radius valley (a paucity of planets around $\sim$1.9 $R_\oplus$) which separates rocky, compact super-Earths and sub-Neptunes with lower bulk densities \citep{2013ApJ...775..105O,2017AJ....154..109F, 2020ApJS..247...28H}.}
Until now, {\jw a number of theoretical models have been proposed to explain the radius valley.} 
{\jw These models can be generally divided into two categories: evolutionary models and primordial models.}
{\jw
On the one hand, from the view of evolutionary models,
the radius valley is a result of the evolution of planetary radius distribution due to the loss of planetary atmosphere after planet formation.
The energy source that drives the atmosphere loss process could be either from outside, i.e., the high energy (e.g., X-ray) radiation of the host star \citep[the photo-evaporation mechanism,][]{2013ApJ...775..105O,2014ApJ...795...65J,2016ApJ...818....4L,2017ApJ...847...29O,2018ApJ...853..163J} or from inside, i.e., the cooling luminosity of central planet core \citep[the core-powered mass loss mechanism,][]{2016ApJ...825...29G,2018MNRAS.476..759G,2019MNRAS.487...24G,2020MNRAS.493..792G}.
}
{\jw 
On the other hand, from the view of primordial models, the radius valley is a natural result from planet formation and migration. 
Some studies suggested that the valley emerged because of the formation of two distinct planet populations with two different core compositions, i.e.,  super-Earths with rocky cores which probably formed in situ, and sub-Neptunes with water/ice-rich cores which probably formed beyond ice lines but migrated into current orbits \citep{2019PNAS..116.9723Z,2020A&A...643L...1V}.
Alternatively, some other studies found that the valley may be recovered by planets formed in situ with the same core composition if the cores have a broad initial mass function and accrete gas in gas-poor (but not gas-empty) nebula \citep{2016ApJ...817...90L,2021ApJ...908...32L}.  
}


{\jw To better understand the origin of the radius valley and further constrain the above theoretic models, one may rely on more observational clues from the dependence of the radius valley on planetary and stellar properties \citep[e.g.,][]{2021MNRAS.508.5886R}.
Here, we briefly summarize recent progress in this aspect.
}
{\jw 
\begin{itemize}
    \item \emph{Planet period} dependence. The valley center is found to be anti-correlated with orbital period of planets \citep{2017AJ....154..109F,2018MNRAS.479.4786V,2019ApJ...875...29M}.
    \item \emph{Stellar mass} dependence. The valley center is positively correlated with stellar mass \citep{2018AJ....156..264F,2019ApJ...874...91W,2020AJ....160..108B}.
    \item \emph{Stellar metallicity} dependence. \cite{2018MNRAS.480.2206O} found that sub-Neptunes are larger and the radius valley is wider around  metal-richer stars.
    \item \emph{Stellar age} dependence. Both \cite{2020AJ....160..108B} and \cite{2021ApJ...911..117S} found the number ratio of super-Earth to sub-Neptune rises around older stars.
\end{itemize}
}

{\jw 
In addition, \cite{2021AJ....161..265D} reported that the radius valley is emptier around younger stars using the California-Kepler Survey (CKS) sample \citep{2017AJ....154..107P,2017AJ....154..108J}.
However, such a feature is not seen by  \cite{2020AJ....160..108B} using the Gaia-Kepler catalog.
Two potential reasons could cause such inconsistent results. 
First, estimation of stellar age is generally difficult and suffers considerable uncertainty. 
Both \cite{2020AJ....160..108B} and \cite{2021ApJ...911..117S} used isochrone fitting to derive age, whose uncertainty is large for main sequence stars, e.g, $\sim 56\%$ in the Gaia-Kepler catalog \citep{2020AJ....159..280B}.
Second, age is generally correlated to other stellar properties, thus the valley dependence on other stellar properties (e.g., metallicity) could affect the observed (apparent) result of the age dependence.

In this paper, we investigate the radius valley morphology in the Galactic context, focusing on the its dependence on Galactic components (e.g., thin/thick disk stars), metallicity ( e.g., $\rm [Fe/H]$ and $\rm [\alpha/Fe]$) and age.
This is the third paper of the series of Planets Across Space and Time (PAST).
In the first paper of PAST \citep[hereafter PAST \uppercase\expandafter{\romannumeral1};][] {2021ApJ...909..115C}, we extended the applicable range of the kinematic method for classification of Galactic components from the Solar neighborhood ($\sim$100–200 pc) to $\sim1500$ pc to cover most of planet hosts, and refined the Age-Velocity dispersion Relation (AVR) to derive 
kinematic ages of planet hosts with a typical uncertainty of $\sim10\%-20\%$.
Applying the methods of PAST \uppercase\expandafter{\romannumeral1}, we constructed a LAMOST-Gaia-Kepler stellar catalog in the second paper of PAST \citep[hereafter PAST \uppercase\expandafter{\romannumeral2};][] {2021AJ....162..100C}, which provides kinematic properties and other basic stellar properties of 35,835 Kepler stars, including hosts of 1061 Kepler planets (candidates). 
The LAMOST-Gaia-Kepler stellar catalog of PAST \uppercase\expandafter{\romannumeral2} allow us to reveal the effects of various stellar properties ($\rm [Fe/H]$, $\rm [\alpha/Fe]$ and age) on the radius valley, respectively.   
}

{\jw The rest of this paper is organized as follows. 
In section \ref{sec.methods}, we describe the sample selection of stars and planets (Sec.\ref{sec.methods.sample}) and define several metrics to characterize the radius gap morphology (Sec.\ref{sec.methods.statistics}).
In section \ref{sec.res}, we present the results of how the radius gap morphology is affected by Galactic component (Sec.\ref{sec.res.TDD}), kinematic age (Sec.\ref{sec.res.kineage}), $\rm [Fe/H]$ (Sec.\ref{sec.res.FeH}) and $\rm [\alpha/Fe]$ (Sec.\ref{sec.res.alpha}).
In section \ref{sec.dis}, we discuss our results and their implications to planet formation and evolution. 
Finally, we summarize the paper in section \ref{sec.sum}. 
}



\section{Methods}
\label{sec.methods}
\subsection{Sample Selection: Stars and Planets}
\label{sec.methods.sample}
{\jw We initialized our stellar sample from the LAMOST-Gaia-Kepler catalog from PAST \uppercase\expandafter{\romannumeral2} \citep{2021AJ....162..100C}, a total of 35835 Kepler stars, including stellar hosts of {\cdc 1,060} Kepler planets (candidates).
In addition to some basic stellar parameters (e.g., mass, radius, effective temperature, $\rm [Fe/H], \ [\alpha/Fe]$), the catalog provides stellar kinematic properties, such as the Galactic velocities and the derived relative membership probabilities among different Galactic components (TD/D, TD/H, Herc/D, Herc/TD), where D, TD, Herc and H denote the thin disk, the thick disk, the Hercules stream and the halo of the Milky Way Galaxy, respectively. }

{\jw For the star sample considered in this paper}, we {\jw first} excluded giant stars by eliminating stars with $\log_{10} {\frac{R_*}{R_\odot}} > 0.00035 \times (T_{\rm eff}-4500)+0.15$ according to \cite{2017AJ....154..109F}.
{\cdc The stellar mass, radius and $T_{\rm eff}$ are adopted from the Gaia-Kepler stellar properties catalog of  \cite{2020AJ....159..280B}. }
{\jw Following \cite{2014A&A...562A..71B}, we then adopted the criteria $``Herc/D<0.5\,\&\, Hec/TD<0.5\,\&\,TD/H>1"$ to keep only stars in the Galactic disk because kinematic age only applies to} stars belonging to the Galactic disk components as suggested in PAST \uppercase\expandafter{\romannumeral1} \citep{2021ApJ...909..115C}.
Finally, we removed stars without $\rm [Fe/H]$ or $\rm [\alpha/Fe]$ measurements.

{\jw For the planet sample, we excluded planets of grazing transits (i.e., $\frac{R_{\rm p}}{R_*} + b > 1$, where {\cdc $b$ is the transit impact parameter in the Kepler DR25 table  from the NASA exoplanet archive \citep{koidr25} and $R_{\rm p}$ is calculated by multiplying stellar radius and the radius ratio of planet to star}), because their radius measurements generally suffer from large uncertainties.}
To ensure the detection efficiency of super-Earths, we only {\jw kept} planets with orbital period less than 100 days.
We also excluded ultra-short-period planets (USPs, planets with orbital period less than 1 day) {\jw because they are relatively rare and likely to be formed differently as compared to the bulk of Kepler planets}  \citep{2018NewAR..83...37W}.
{\jw Further, since we mainly focus on planets close to the radius valley, thus we only kept planets with radii in the range of $1-6 R_\oplus$.}
After the above selections, we are left with 446 stars hosting 621 planets (candidates).
{\jw Table \ref{tab:sampleprocedure} is a rundown of the above sample selection process.}

\begin{figure}[!t]
\centering
\includegraphics[width=0.75\textwidth]{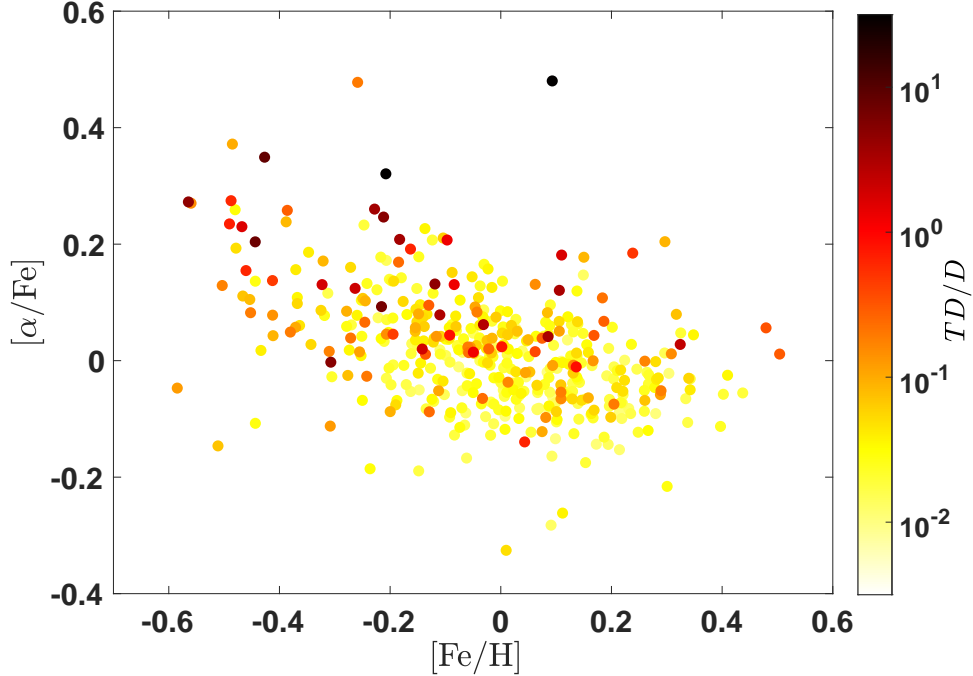}
\caption{The $\rm [Fe/H]-[\alpha/Fe]$ diagram color-coded by the relative probability between thick disk to thin disk $TD/D$ for the stellar sample (section \ref{sec.methods.sample}).
\label{figFeHalphaTDDwholesample}}
\end{figure}

Figure \ref{figFeHalphaTDDwholesample} displays the color-coded distributions of relative probabilities between thick disk to thin disk ($TD/D$) in the $\rm [Fe/H]-\rm [\alpha/Fe]$ plane for our stellar sample. 
As can be seen, in general, stars with larger $TD/D$ (kinematically thicker) have lower $\rm [Fe/H]$ and higher $\rm [\alpha/Fe]$, which {\jw is consistent} with previous studies \citep[e.g.,][]{2014A&A...562A..71B,2021ApJ...909..115C}.

\begin{table}[!t]
\renewcommand\arraystretch{1.35}
\centering
\caption{{\jw Rundown of the Sample Selection}}
{\footnotesize
\label{tab:sampleprocedure}
\linespread{1.8}
\begin{tabular}{l|cc} \hline
Selection criteria & $N_{\rm s}$ & $N_{\rm p}$ \\\hline
LAMOST-Gaia-Kepler sample (PAST \uppercase\expandafter{\romannumeral2}) & 764 & 1060 \\
Excluding giant stars  & 747 & 1040 \\
Belong to Galactic disks & 659 & 919 \\
With $\rm [Fe/H]$ and $\rm [\alpha/Fe]$ measurements & 605 & 850 \\
$\frac{R_{\rm p}}{R_*} + b \leq 1$ &  589 & 833 \\
Period $\le 100$ days & 544 & 767 \\ 
Period $\ge 1$ days & 532 & 753 \\ 
$1 R_\oplus <R_{\rm p}< 6 R_\oplus$ & 446 & 621 \\  \hline
\end{tabular}}
\flushleft
{\scriptsize
 $N_{\rm s}$ and $N_{\rm p}$ are the numbers of host stars and planets during the process of sample selection in section \ref{sec.methods.sample}.}
\end{table}

\begin{figure}[!t]
\centering
\includegraphics[width=0.75\textwidth]{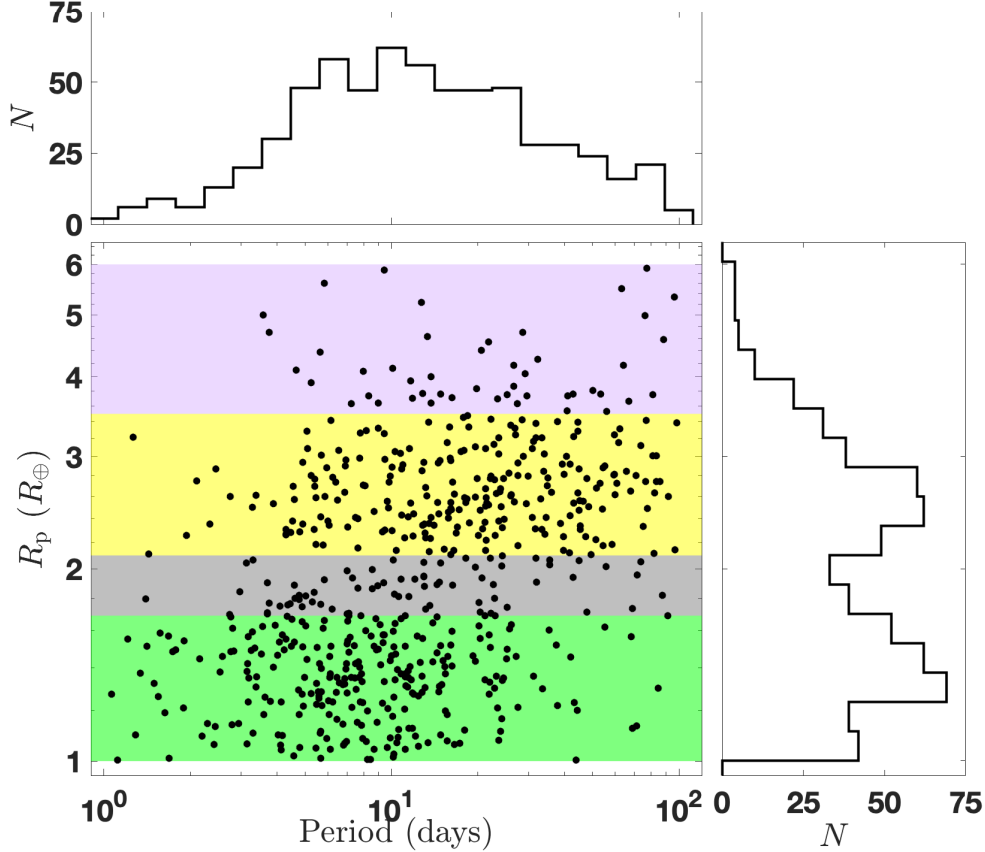}
\caption{The period-radius diagram for the planetary sample (section \ref{sec.methods.sample}).
Areas of different colors represent planets of different radius: green for super-Earths ($1-1.7 R_\oplus$), grey for valley planets ($1.7-2.1 R_\oplus$), yellow for sub-Neptunes ($2.1-3.5 R_\oplus$) and purple for Neptunes ($3.5-6 R_\oplus$).
Histograms of planetary radius ($R_{\rm p}$) and orbital period ($P$) are shown in the right panel and the top panel respectively.
\label{figPRwholesample}}
\end{figure}

According to previous studies \citep[e.g.,][]{2013ApJ...775..105O,2017ApJ...847...29O,2017AJ....154..109F,2018AJ....156..264F,2021arXiv210302127Z}, the radius valley {\jw is} located at $\sim 1.9 \pm 0.2 R_{\oplus}$. 
We then divide {\jw our planet sample} into four sub-samples according to their {\jw radii}:
{\jw 
\begin{enumerate}
    \item Valley planet (VP): $1.7-2.1 R_{\oplus}$ ($N_{\rm VP}$=70);
    \item Super-Earth (SE): $1.0-1.7 R_{\oplus}$ ($N_{\rm SE}$=266);
    \item Sub-Neptune (SN): $2.1-3.5 R_{\oplus}$ ($N_{\rm SN}$=238);
    \item Neptune-size planet (NP): $3.5-6 R_\oplus$ ($N_{\rm NP}$=47).
\end{enumerate}
Here, $N_{\rm VP}$, $N_{\rm SE}$, $N_{\rm SN}$ and $N_{\rm NP}$ are the numbers of planets in the corresponding sub-samples, respectively.
}

Figure \ref{figPRwholesample} shows the period-radius distributions of our planetary sample. 
As can be seen, there exists an obvious bimodal distribution in planetary radius and a valley in $\sim ~1.9 R_\oplus$, which is well consistent with previous studies \citep[e.g,][]{2017AJ....154..109F}.
In the subsequent {\jw sections}, we will {\jw further characterize the morphology of the radius valley and explore its dependence on various stellar properties.}

\subsection{Characterizing the Radius Valley Morphology}
\label{sec.methods.statistics}
{\jw To characterize the morphology of the radius valley, we adopted a set of metrics, which are defined as follows.}
\begin{enumerate}
    \item  {\jw The contrast of radius valley $C_{\rm valley}$, defined as the number ratio of super-Earths plus sub-Neptunes to the valley planets, i.e.,
    \begin{equation}
    C_{\rm valley} = \frac{N_{\rm SE}+N_{\rm SN}}{N_{\rm VP}}.
    \label{eqDV}
    \end{equation} }
    \item  {\jw The asymmetry on the two sides of the valley defined as the number ratio (in logarithm) of super-Earths to sub-Neptunes, i.e., \begin{equation}
    A_{\rm valley} = \log_{10} \left( \frac{N_{\rm SE}}{N_{\rm SN}} \right).
    \label{eqREN}  
    \end{equation} }
    \item {\jw The average radius of planets with size larger than Valley planets, i.e., 
    \begin{equation}
    {R}_{\rm valley}^{+} = \frac{1}{N_{\rm SN}+N_{\rm NP}}\sum^{N_{\rm SN}+N_{\rm NP}}_{i}{R_{\rm i}}, 
    \label{eqRNSE}  
    \end{equation}
    where $2.1R_\oplus <R_{\rm i}<6R_\oplus$. }
    \item {\jw The average radius of planets with size smaller than Valley planets, i.e., 
    \begin{equation}
    {R}_{\rm valley}^{-} = \frac{1}{N_{\rm SE}}\sum^{N_{\rm SE}}_{i}{R_{\rm i}}, 
    \label{eqRSE}  
    \end{equation}
    where $1.0R_\oplus <R_{\rm i}<1.7 R_\oplus$. }
    \item {\jw The number fraction of Neptune-size planets in the whole planet sample , i.e.,
    \begin{equation}
    f_{\rm NP} = \frac{N_{\rm NP}}{N_{\rm SE}+N_{\rm VP}+N_{\rm SN}+N_{\rm NP}}.
    \label{eqfNep}  
    \end{equation}
    This metric reflects the extension degree of the second (with larger radius) peak beside the radius valley.}

\end{enumerate}

To obtain the uncertainties of these metrics, following \cite{2020AJ....160..108B}, we {\jw performed} Monte Carlo simulations by resampling the planet {\jw radii from a normal distribution given their measured values and uncertainties.}
{\jw We then re-counted $N_{\rm VP}$, $N_{\rm SE}$, $N_{\rm SN}$ and $N_{\rm NP}$ and computed those metrics, i.e., $A_{\rm valley}$, $C_{\rm valley}$,  ${R}_{\rm valley}^{+}$, ${R}_{\rm valley}^{-}$ and $f_{\rm NP}$ for the resampled data.}
We repeated the {\jw above procedure} for 10,000 times, then {\jw calculated the 50$\pm$34.1 percentiles in the resampled distributions as the uncertainties of the corresponding metrics.}

\section{Results}
\label{sec.res}
{\jw In this section, we explore how the radius valley morphology (Eq. (1)$-$(5)) depends on Galactic component (e.g., thin/thick disk, Sec.\ref{sec.res.TDD}), kinematic age  (Sec. \ref{sec.res.kineage}), $\rm [Fe/H]$ (Sec. \ref{sec.res.FeH}) and $\rm [\alpha/Fe]$ (Sec. \ref{sec.res.alpha}).
}
\begin{figure*}[!t]
\centering
\includegraphics[width=\textwidth]{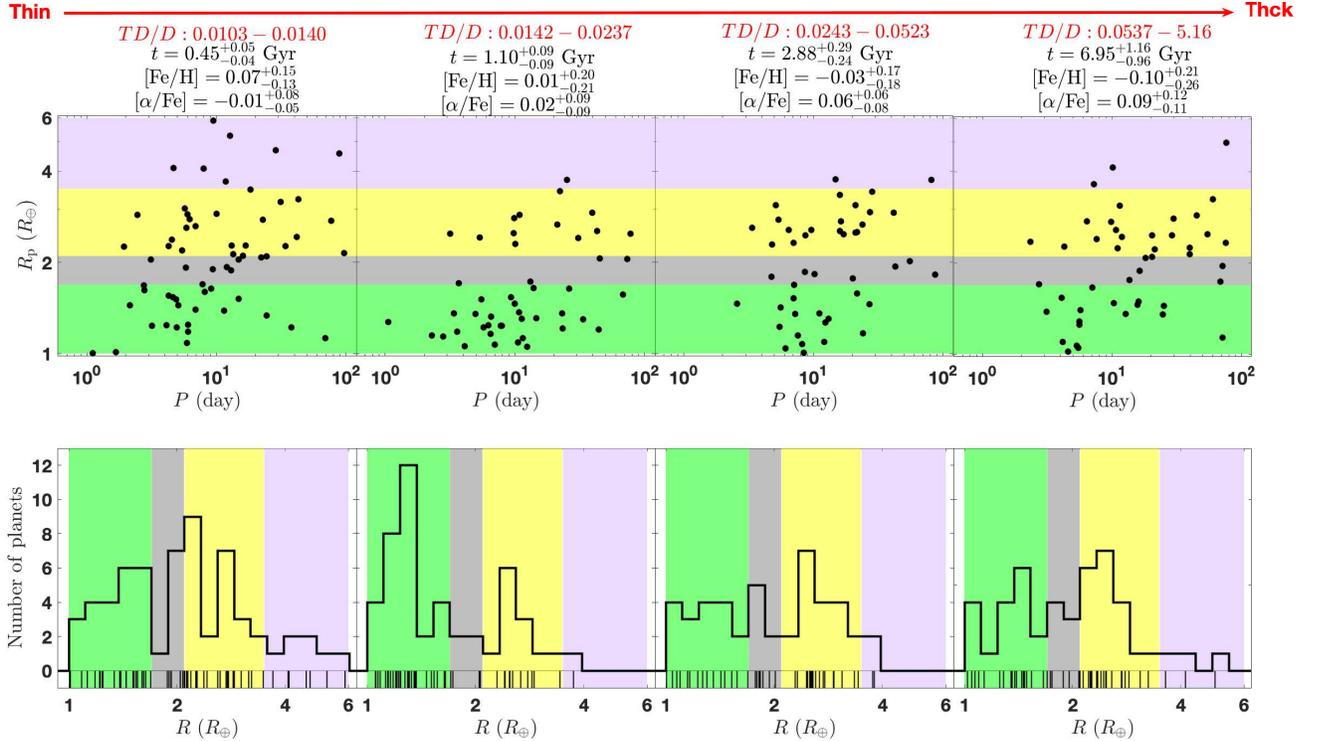}
\caption{The period-radius diagram (upper panel) and radius distribution (bottom panel) of planets in different $TD/D$ bins from our sample based on the LAMOST-Gaia-Kepler catalog.  Different colors represent planets of different radius: green for super-Earths ($1-1.7 R_\oplus$), grey for valley planets ($1.7-2.1 R_\oplus$), yellow for sub-Neptunes ($2.1-3.5 R_\oplus$) and purple for Neptune-size planets ($3.5-6 R_\oplus$). {\jw The median values and $1-\sigma$ uncertainties of kinematic age, $\rm [Fe/H]$ and $\rm [\alpha/Fe]$ of each bin are printed at the top.}
\label{figRadiusValleyTDD}}
\end{figure*}

\subsection{Dependence of Radius Valley on Galactic {\jw Component}}
\label{sec.res.TDD}
{\jw In this paper, we only considered the Galactic disk components, i.e., thin and thick disks, and used the relative probability, $\rm TD/D$ to quantify the likelihood that a star belong to the thick disk relative to the thin disk.}
The higher the $TD/D$ is, the star is more likely to belong to the thick disk than thin disk.
{\jw We investigated the radius valley morphology as a function of $TD/D$.
To do this, we divided our sample into four bins according to $TD/D$.
Specifically, we first sorted the whole sample in the order in which TD/D increases. 
We took 10\% (44) of stars at the lower $\rm TD/D$ end as the first bin, and divided the rest into three bins of equal size ($134$) according to their $\rm TD/D$.
To remove the effects of other stellar parameters and planet detection efficiency on different bins, the latter three bins were further controlled to let them have similar distributions in stellar mass, radius and photometric precision (i.e., Combined Differential Photometric Precision, CDPP) as compared to the first bin.
This was realized by adopting the NearestNeighbors method in the scikit-learn \citep{10.5555/1953048.2078195} to select the nearest neighbor in the space of the controlled parameters from stars in the latter three bins for every star in the first bin (see the Appendix for the details of the construction and validation of the control bins).
After the above parameter control process, the latter three bins have similar distributions in stellar mass, radius and CDPP as compared to the first bin (Kolmogorov-Smirnov (KS) test P value greater than 0.8 as shown in Fig \ref{figCDFMRCDPPTDD}).
{\CDC We stress that, this method removes the difference in sample completeness between bins, but it does not derive the intrinsic distribution in each bin by directly applying completeness corrections, so our statistical results are meaningful in the differential sense among various bins.}
Note, here we did not control $[\rm Fe/H]$ and $\rm [\alpha/Fe]$, because the differences in stellar metallicity between thin and thick disk stars are essential and inevitable. 
}

Figure \ref{figRadiusValleyTDD} displays the period-radius diagram (upper panel) and the radius distribution (bottom panel) of planets in different $TD/D$ bins {\jw (after parameter control)}.
{\jw As expected and printed at the top of the figure, with the increase of $TD/D$ (i.e., from thin disk to thick disk), the kinematic age and $\rm [\alpha/Fe]$ increase, while the $\rm [Fe/H]$ decreases.
However, there seems no apparent pattern between radius valley and TD/D.
Such an intuition is confirmed in Figure 4, in which we plot the five metrics ($C_{\rm valley}$, $A_{\rm valley}$, ${R}^{+}_{\rm valley}$,  ${R}^{-}_{\rm valley}$ and $f_{\rm NP}$, Equations (1)-(5)) of radius valley morphology as a function of $TD/D$.
As can be seen, with the increase of $ TD/D$, $C_{\rm valley}$ and $A_{\rm valley}$ seem to first increase then decrease, while  ${R}_{\rm valley}^{-}$ and $f_{\rm NP}$ first decrease then increase.
{\CDC We have also adjusted the size of the first bin (e.g., 15\%) and found that the evolution of the radius valley and the five metrics generally maintains, demonstrating our results are not (significantly) dependent on the selection of first bins.}

{\cdc However, there is no clear monotonic trend between $TD/D$ and any of the morphology metrics due to such small number statistics.}
Such a non-monotonic result is not unexpected because thick disk stars are  intrinsically older with lower (higher) $\rm [Fe/H]$ ($\rm [\alpha/Fe]$) compared to thin disk stars. 
Thus, the results of increasing $TD/D$ are essentially a combination of the effects of growing age, decreasing $\rm [Fe/H]$ and increasing $\rm [\alpha/Fe]$.
In the following subsections, we will investigate these effects separately.
}

\begin{figure}[!t]
\centering
\includegraphics[width=0.5\textwidth]{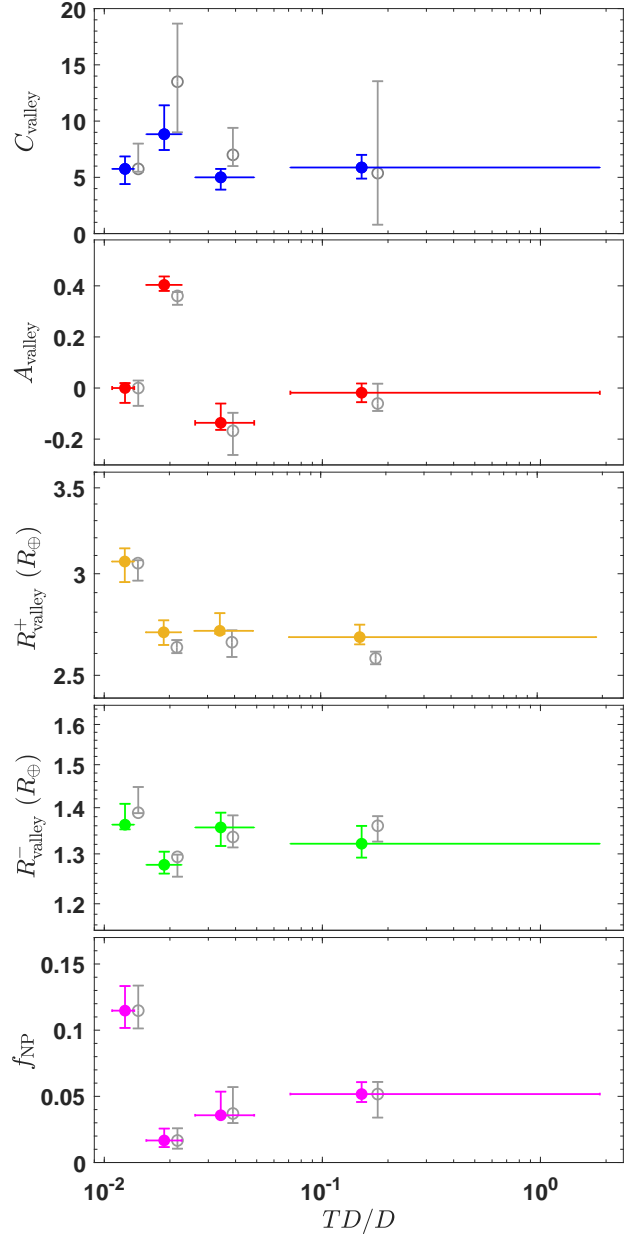}
\caption{The five statistics to characterize the radius valley morphology (i.e., $C_{\rm valley}$, $A_{\rm valley}$, ${R}^{+}_{\rm valley}$, ${R}^{-}_{\rm valley}$, and $f_{\rm NP}$) as functions of the relative probability between thick disk to thin disk (TD/D).
{\cdc In section \ref{sec.dis.mp}, we rescale the planet radii to eliminate the influence of stellar mass and planetary orbital period on the boundary of radius valley. The results with rescaling radii are plotted as open circles and shifted to the right to avoid overlapping.}
\label{DvalleyRENTDD}}
\end{figure}

\subsection{\jw Dependence of Radius Valley on Age}
\label{sec.res.kineage}

\begin{figure*}[!t]
\centering
\includegraphics[width=\textwidth]{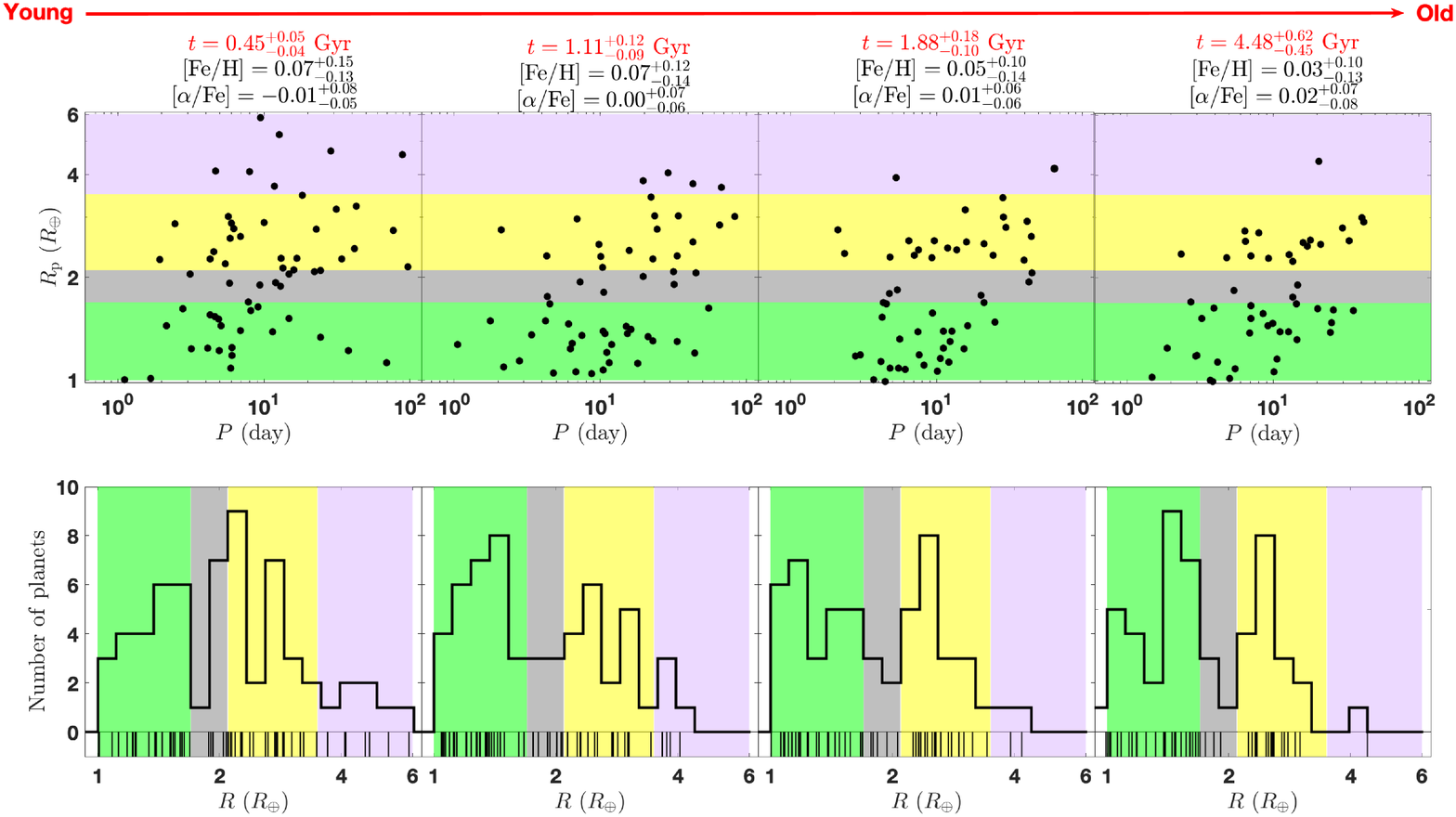}
\caption{The period-radius diagram (upper panel) and radius distribution (bottom panel) of planets with  different kinematic ages $t \ (\rm Gyr)$ from our selected sample from LAMOST-Gaia-Kepler catalog.
The median values and $1-\sigma$ uncertainties of kinematic age, $\rm [Fe/H]$ and $\rm [\alpha/Fe]$ of each bin are printed at the top..
\label{figRadiusValleyAge}}
\end{figure*}

In this subsection, we focus on the effect of age on the radius valley morphology.
{\cdc Similar to the parameter control method as in section \ref{sec.res.TDD}, we further control $\rm [Fe/H]$ and $\rm [\alpha/Fe]$ to isolate the effect of age as shown in Fig \ref{figCDFMRCDPPAge}.}

\begin{figure}[!t]
\centering
\includegraphics[width=0.5\textwidth]{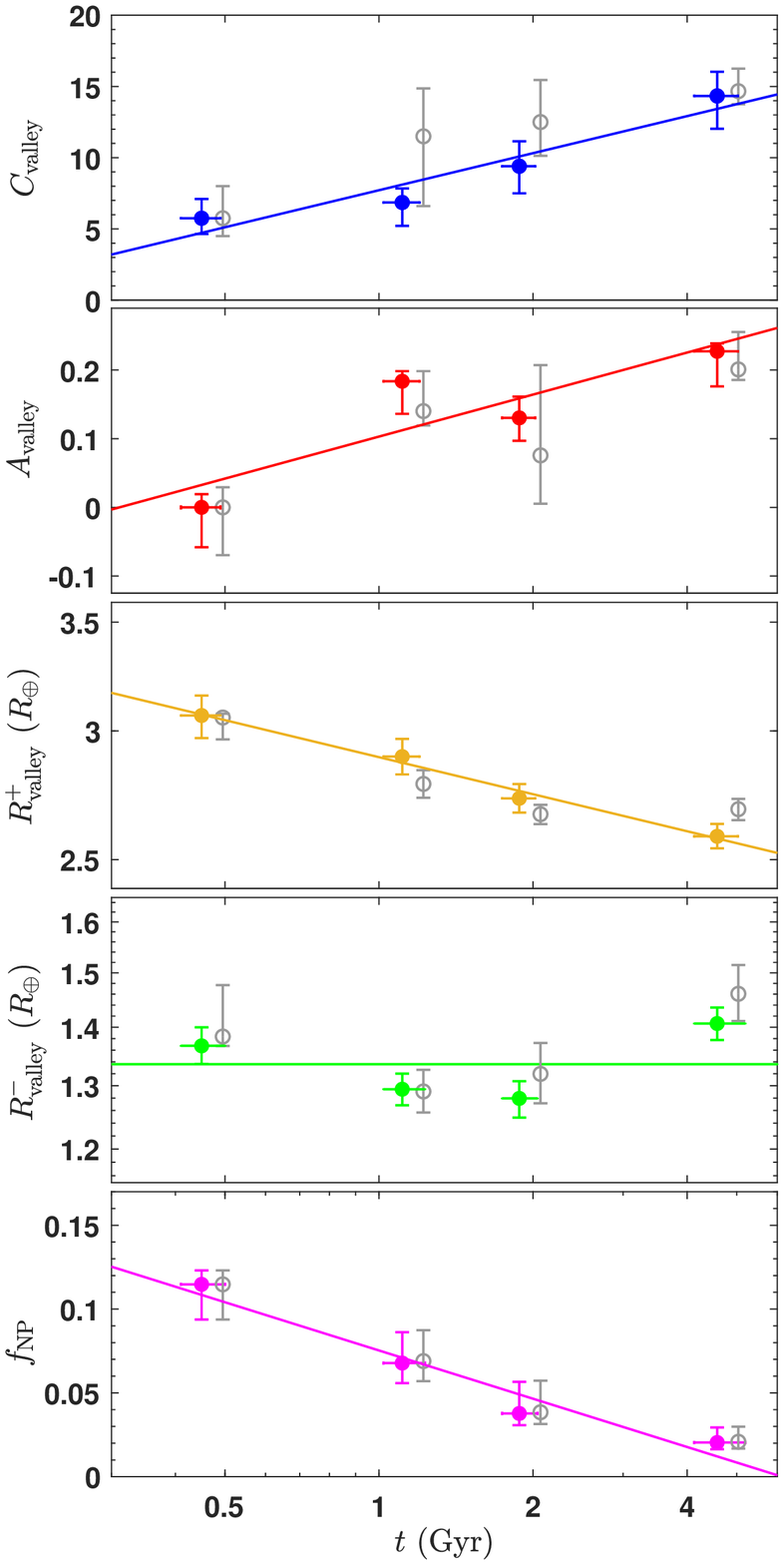}
\caption{The five statistics to characterize the radius valley morphology (i.e., $C_{\rm valley}$, $A_{\rm valley}$, ${R}^{+}_{\rm valley}$, ${R}^{-}_{\rm valley}$, and $f_{\rm NP}$) as functions of kinematic age $t \ (\rm Gyr)$.
{\cdc In section \ref{sec.dis.mp}, we rescale the planet radii to eliminate the influence of stellar mass and planetary orbital period on the boundary of radius valley. The results with rescaling radii are plotted as open circles and shifted to the right to avoid overlapping.}
\label{DvalleyRENAge}}
\end{figure}

{\jw We estimated the average age of each bin with the kinematic method by using the Age-Velocity dispersion Relation (AVR) that refined by PAST-I \citep{2021ApJ...909..115C}. }
In Figure \ref{figRadiusValleyAge}, we show the period-radius diagram (upper panel) and radius distribution (bottom panel) of planets in each bin.
The median values and $1-\sigma$ uncertainties of kinematic age, $\rm [Fe/H]$ and $\rm [\alpha/Fe]$ of each bin are printed at the top.
{\jw From left to right, the age increases monotonically and significantly while $\rm [Fe/H]$ and $\rm [\alpha/Fe]$ are almost unchanged among different bins, which is expected after the above parameter control.}
{\jw Some apparent age trends emerge in Figure  \ref{figRadiusValleyAge}.
As can be seen, the radius valley looks more prominent in older bins, and the number ratio of super-Earths to sub-Neptunes seems to increase with age.
In addition, the radius range of planets above the radius valley is shrinking over the age.

We further investigate the above apparent trends quantitatively by calculating the five radius morphology metrics (Equtions (1)-(5)) in each bin. 
Figure \ref{DvalleyRENAge} shows these metrics, i.e., $C_{\rm valley}$, $A_{\rm valley}$, ${R}^{+}_{\rm valley}$,  ${R}^{-}_{\rm valley}$ and $f_{\rm NP}$, as a function of age from the top to bottom panels. 
For each panel in Figure \ref{DvalleyRENAge}, we fit the data points with two models, i.e, a constant model ($\rm y=C$) and a linear model with a logarithm time scale ($y=A\times \log_{10}(t)+B$), or in log-log space ($ \log_{10}(y)=A\times \log_{10}(t)+B$).
We adopted two approaches to evaluate the two models.
{\cdc First, the simplest one is to calculate the Akaike information criterion (AIC) score of the constant and linear models, $\rm AIC_{con}$ and $\rm AIC_{lin}$.
In term of the residual sum of squares (RSS), the AIC can be calculated with the following formula
\citep{CAVANAUGH1997201}:
\begin{eqnarray}
{\rm AIC} = 2k+n \ln({\rm RSS}) , 
\label{AIC}
\end{eqnarray}
where $n$ is the number of data points, $k$ is the number of model parameters.
We then calculate the difference in the AIC scores ($\Delta \rm AIC \equiv AIC_{con}-AIC_{lin}$) of the two models and the model with smaller AIC are preferred.}
Second, we adopted a Monte Carlo simulation by resampling the data points given their uncertainty and refitting the data for 10,000 times. 
The confidence level of the best-fit parameter is calculated as the fraction of these resample-fit trials in which the model has a lower AIC score.
The fitting results are as follows:

\begin{enumerate}
    \item For $C_{\rm valley}$, the linear increasing model is preferred than the constant model with a $\rm \Delta  AIC=7.65$ and a confidence level of 95.21\% from Monte Carlo simulation. 
    The best-fit is 
    {\cdc \begin{equation}
    C_{\rm valley} = 8.6^{+2.5}_{-3.4} \times \log_{10}  t+ 7.7^{+0.4}_{-1.5}.
    \label{eqC_Age}
    \end{equation}}

    \item For $A_{\rm valley}$, the linear increasing model is preferred than the constant model with a $\rm \Delta  AIC=5.66$ and a confidence level of 96.57\% from Monte Carlo simulation.  
    The best-fit is 
    {\cdc \begin{equation}
    A_{\rm valley} = 0.20^{+0.08}_{-0.03} \times \log_{10} t+ 0.10^{+0.01}_{-0.03}.
    \label{eqA_Age}
    \end{equation}}
   Although we simply fit a linear function here due to limited data points, we note that the increase of $A_{\rm valley}$ (or the number ratio of super-Earths to sub-Neptunes equivalently) seems nonlinear; it is mainly completed within $\sim1-2$  Gyr (see more discussions in 4.2.1)  

    \item For ${R}^{+}_{\rm valley}$, the linear decay model is preferred than the constant model with a $\rm \Delta  AIC=12.60$ and a confidence level of 99.71\% from Monte Carlo simulation.
    The best-fit is 
    {\cdc \begin{equation}
    \log_{10}\left(\frac{{R}^{+}_{\rm valley}}{R_\oplus}\right) = -0.08^{+0.01}_{-0.01} \times \log_{10} t+ 0.46^{+0.04}_{-0.05}.
    \label{eqR+_Age}
    \end{equation}}

    \item For ${R}^{-}_{\rm valley}$, the constant model is preferred than the linear model with a $\rm \Delta  AIC=2.13$ and a confidence level of 99.32\% from Monte Carlo simulation.  
   {\cdc The best-fit is 
    \begin{equation}
        {R}^{-}_{\rm valley}= 1.34^{+0.04}_{-0.16} R_\oplus.
    \label{eqR-_Age}
    \end{equation}}

    \item For $f_{\rm NP}$, the linear decreasing model is preferred than the constant model with a $\rm \Delta  AIC=10.12$ and a confidence level of 99.85\% from Monte Carlo simulation. 
    The best-fit is 
    {\cdc \begin{equation}
    f_{\rm NP} = -0.10^{+0.01}_{-0.02} \times \log_{10} t+ 0.08^{+0.02}_{-0.02}.
    \label{eqfNP_Age}
    \end{equation}}
\end{enumerate}
}

{\jw Above quantitative results confirm the apparent trends seen in Figure \ref{figRadiusValleyAge}.
With increasing age over Gyr timescales, the radius valley become more prominent (i.e., larger $C_{\rm valley}$) and more asymmetric (larger $A_{\rm valley}$ or larger number ratio of super-Earths to sub-Neptunes); the average radii of planets larger than valley planets decreases (with a slope of ${\rm d} (\log {R}^{+}_{\rm valley})/{\rm d} (\log t) \sim -0.07$) while the average radii of planets smaller than valley planets (i.e., super-Earth) changes little; and the fraction of Neptune-size planets $f_{\rm NP}$ decreases significantly.
The implications of these results on planet formation and evolution will be discussed in section \ref{sec.dis.impl}.
}

\subsection{\jw Dependence of Radius Valley on $\rm [Fe/H]$}
\label{sec.res.FeH}

\begin{figure*}[!t]
\centering
\includegraphics[width=\textwidth]{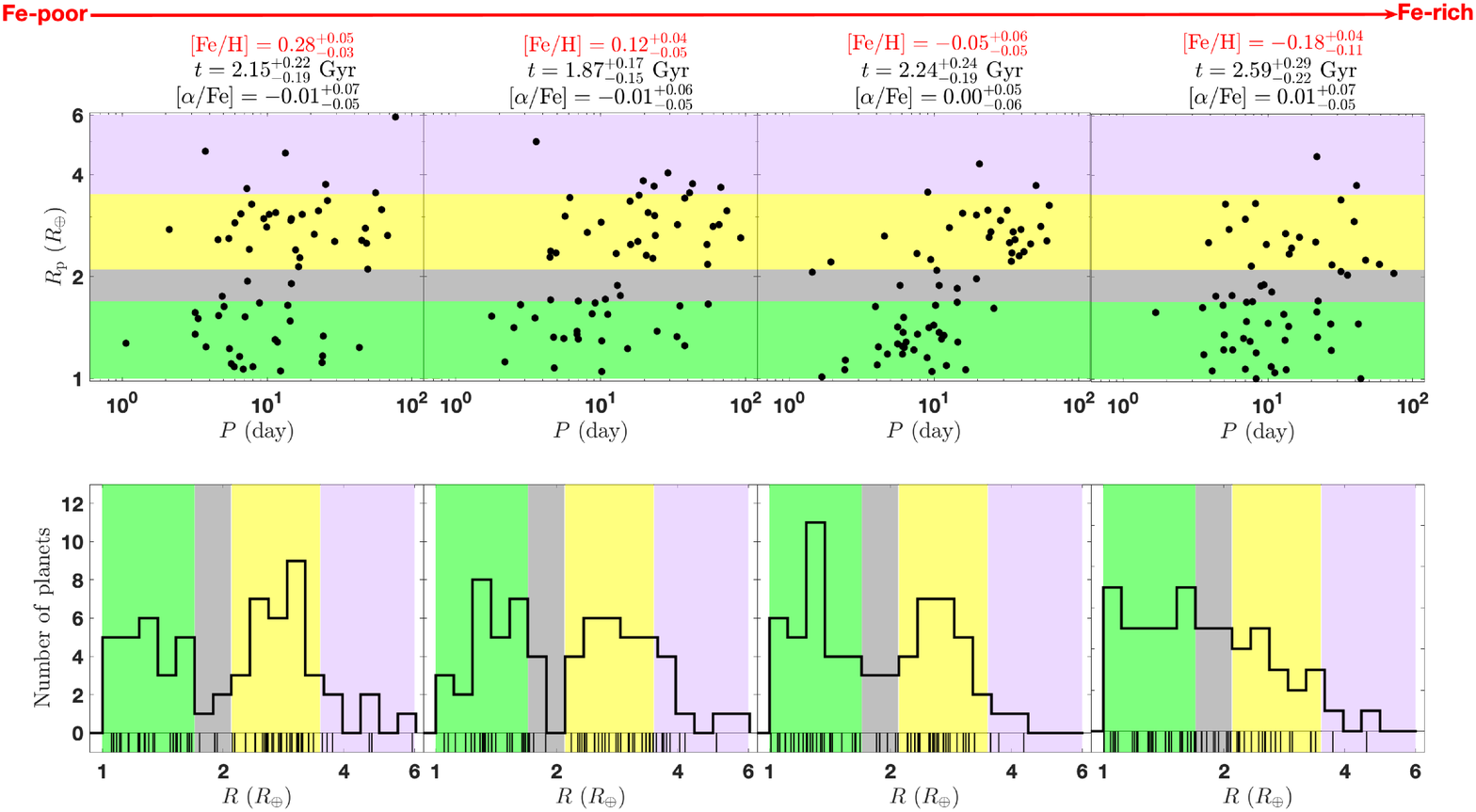}
\caption{The period-radius diagram (upper panel) and radius distribution (bottom panel) of planets with different metallicity $\rm [Fe/H]$ from our selected sample from LAMOST-Gaia-Kepler catalog.
The median values and $1-\sigma$ uncertainties of kinematic age, $\rm [Fe/H]$ and $\rm [\alpha/Fe]$ of each bin are printed at the top.
\label{figRadiusValleyFeH}}
\end{figure*}

{\jw In this subsection, we explore the dependence of radius valley on $\rm [Fe/H]$.
{\cdc Similar to the parameter control method as in sections 3.1, we controlled the parameters, i.e., stellar mass, radius, CDPP, $\rm [\alpha/Fe]$ and TD/D to isolate the effect of $\rm [Fe/H]$.}
After the parameter control process, stars in different bins have similar distributions in these controlled parameters (with all KS test p-values larger than 0.2) as shown in Figure \ref{figCDFMRCDPPFeH}. 
Note, we were not allowed to control stellar age directly. 
In fact, age has been controlled (see the ages printed at the top of Figure \ref{figRadiusValleyFeH})  indirectly by controlling TD/D, because TD/D and age are strongly correlated to each other \citep{2021ApJ...909..115C}.
}

In Figure \ref{figRadiusValleyFeH}, we compare the period-radius diagram (upper panel) and radius distribution (bottom panel) of planets in different $\rm [Fe/H]$ bins after parameter control.
As printed at the top, the median values of $\rm [Fe/H]$ decreases continuously while the kinematic age and $\rm [\alpha/Fe]$ are nearly unchanged {\jw within their uncertainty, demonstrating that these parameters have been well controlled.}
Some apparent trends emerge in Figure \ref{figRadiusValleyFeH}.
{\jw 
As can be seen, the radius valley seems to be filled up as $\rm [Fe/H]$ decreases.
There seems to be more sub-Neptunes relative to super-Earths in the Fe$-$rich bins than in the Fe$-$poor bins.
In addition, the radius range of planets above the radius valley is shrinking from Fe$-$rich bins to Fe$-$poor bins.
}

\begin{figure}[!t]
\centering
\includegraphics[width=0.5\textwidth]{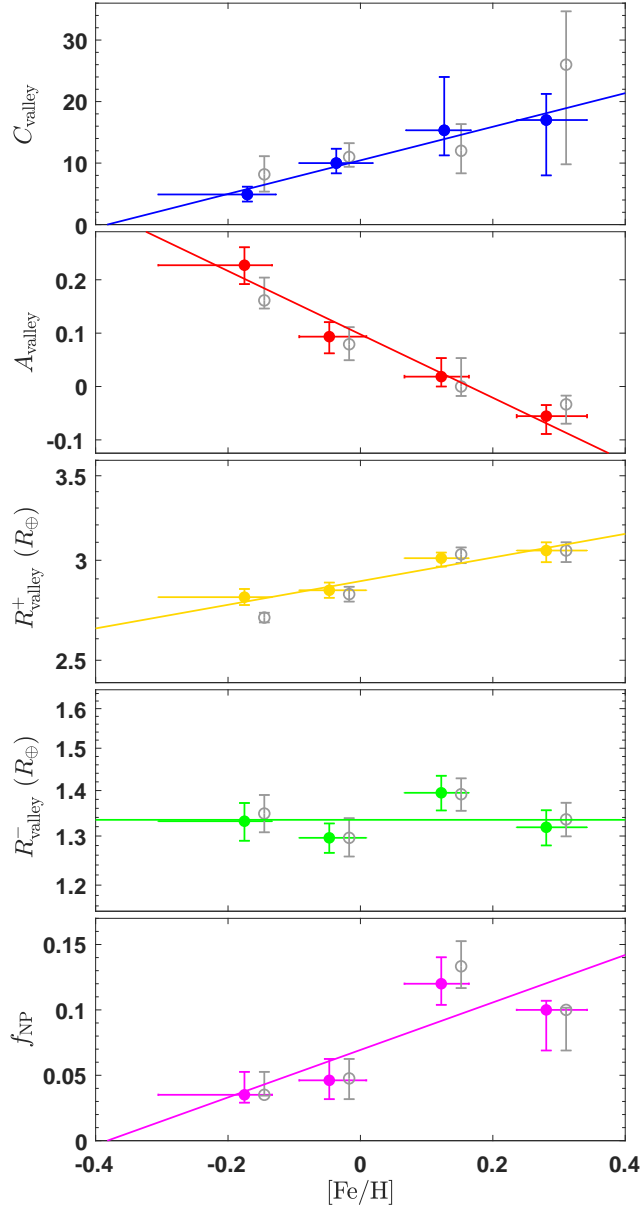}
\caption{The five statistics to characterize the radius valley morphology (i.e., $C_{\rm valley}$, $A_{\rm valley}$, ${R}^{+}_{\rm valley}$, ${R}^{-}_{\rm valley}$, and $f_{\rm NP}$) as functions of  $\rm [Fe/H]$.
{\cdc In section \ref{sec.dis.mp}, we rescale the planet radii to eliminate the influence of stellar mass and planetary orbital period on the boundary of radius valley. The results with rescaling radii are plotted as open circles and shifted to the right to avoid overlapping.}
\label{DvalleyRENFeH}}
\end{figure}

To quantify the above trends, we further calculated the five radius valley morphology (Equation (1)-(5)) for the four bins of different metallicity $\rm [Fe/H]$.
In Figure \ref{DvalleyRENFeH}, we show these metrics, i.e., $C_{\rm valley}$, $A_{\rm valley}$, ${R}^{+}_{\rm valley}$,  ${R}^{-}_{\rm valley}$ and $f_{\rm NP}$, as a function of $\rm [Fe/H]$ from the top to bottom panels. 
We fit the data points in each panel with two models, i.e, a constant model ($\rm y=C$) and a linear model ($y=A\times {\rm [Fe/H]} +B$), or in a logarithm ordinate scale $ \log_{10}(y)=A\times {\rm [Fe/H]}+B$).
{\jw
Then, following the same procedures in section \ref{sec.res.kineage}, we evaluated these two models by calculating the difference in AIC score ($\rm \Delta AIC$) and the confidence level from Monte Carlo simulation.
}
The fitting results are as follows:
\begin{enumerate}
    \item For $C_{\rm valley}$, the linear increasing model is preferred than the constant model with a $\rm \Delta  AIC=10.13$ and a confidence level of 95.72\% from Monte Carlo simulation. 
    The best-fit is 
    {\cdc \begin{equation}
    C_{\rm valley}= 26.3^{+3.4}_{-8.8} \times {\rm [Fe/H]} + 10.4^{+1.5}_{-1.1}.
    \label{eqC_Fe}
    \end{equation}}
    
    \item For $A_{\rm valley}$, the linear decay model is preferred than the constant model with a $\rm \Delta  AIC=10.51$ and a confidence level of 99.92\% from Monte Carlo simulation.  
    The best-fit is 
    {\cdc \begin{equation}
    A_{\rm valley}= -0.6^{+0.1}_{-0.1} \times {\rm [Fe/H]}+ 0.10^{+0.02}_{-0.02}.
    \label{eqA_Fe}
    \end{equation}}
    
    \item For ${R}^{+}_{\rm valley}$, the linear increasing model is preferred than the constant model with a $\rm \Delta  AIC=10.70$ and a confidence level of 99.46\% from Monte Carlo simulation.
    The best-fit is 
    {\cdc \begin{equation}
    \log_{10}\left(\frac{{R}^{+}_{\rm valley}}{R_\oplus}\right)= 0.09^{+0.02}_{-0.03} \times {\rm [Fe/H]} + 0.46^{+0.06}_{-0.02}.
    \label{eqR+_Fe}
    \end{equation}}
    
    \item For ${R}^{-}_{\rm valley}$, the constant model is preferred than the linear model with a $\rm \Delta  AIC=1.87$ and a confidence level of 97.28\% from Monte Carlo simulation.  
    The best-fit is 
    \begin{equation}
        {R}^{-}_{\rm valley}= 1.33^{+0.05}_{-0.04} R_\oplus.
    \label{eqR-_Fe}
    \end{equation}
     
    \item For $f_{\rm NP}$, the linear increasing model is preferred than the constant model with a $\rm \Delta  AIC=4.71$ and a confidence level of 95.23\% from Monte Carlo simulation. 
    The best-fit is 
    {\cdc \begin{equation}
    f_{\rm NP} = 0.18^{+0.04}_{-0.05} \times {\rm [Fe/H]}+0.07^{+0.02}_{-0.01}.
    \label{eqfNp_Fe}
    \end{equation}}
     
\end{enumerate}

{\jw Above quantitative results confirm the apparent trends seen in Figure \ref{figRadiusValleyFeH}.
With the increase of $\rm [Fe/H]$ 
the radius valley become emptier; the ratio of super-Earths to sub-Neptunes decreases significantly; the average radii of planets larger than valley planets increases continuously with a slope of ${\rm d} \log {R}^{+}_{\rm valley}/{\rm d [\rm Fe/H]}  \sim 0.1$ while the average radius of smaller planets, i.e., super-Earth, is almost unchanged;
and the fraction of Neptune-sized planets increase significantly.
The implications of these results on the formation mechanisms of radius valley  will be discussed in section \ref{sec.dis.impl}.
}

\begin{figure*}[!t]
\centering
\includegraphics[width=\textwidth]{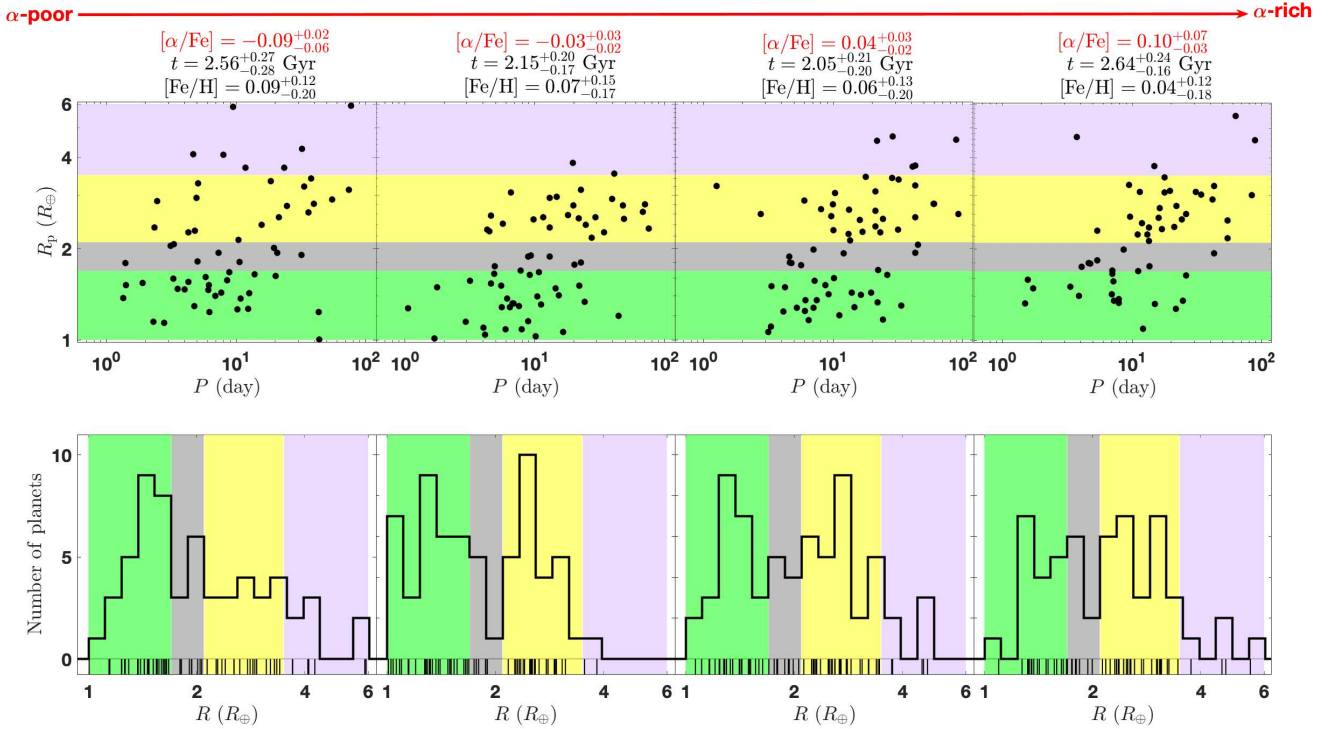}
\caption{The period-radius diagram (upper panel) and radius distribution (bottom panel) of planets with  different $\rm [\alpha/Fe]$ from our selected sample from LAMOST-Gaia-Kepler catalog.
The median values and $1-\sigma$ uncertainties of kinematic age, $\rm [Fe/H]$ and $\rm [\alpha/Fe]$ of each bin are printed at the top.
\label{figRadiusValleyalpha}}
\end{figure*}


\subsection{{\jw Dependence of Radius Valley on $\rm [\alpha/Fe]$}}
\label{sec.res.alpha}
{\jw
In this subsection, we study the link between the radius valley and $\rm [\alpha/Fe]$.
{\cdc Similar to the parameter control method as in section 3.1, we controlled the parameters, i.e., stellar mass, radius, CDPP, $\rm [Fe/H]$ and TD/D to isolate the effect of $\rm [\alpha/Fe]$.}
After the parameter control process, stars in different bins have similar distributions in these controlled parameters (with all KS test p-values larger than 0.15) as shown in Figure \ref{figCDFMRCDPPalpha}. 
}

\begin{figure}[!t]
\centering
\includegraphics[width=0.5\textwidth]{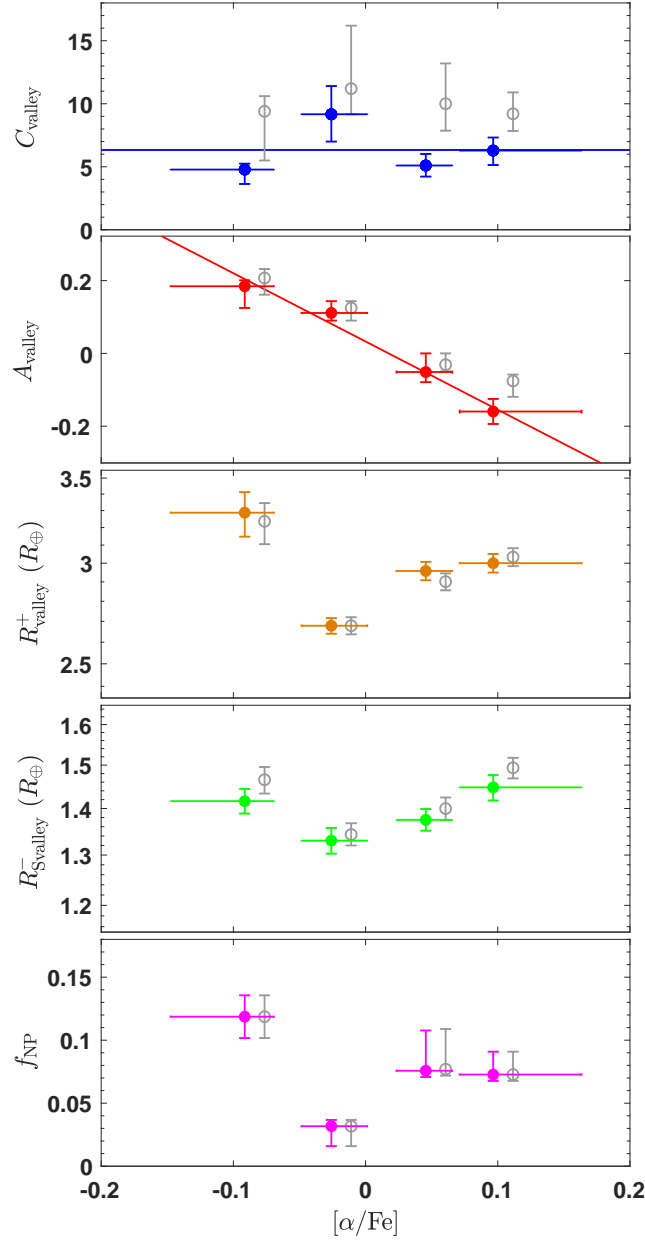}
\caption{The five statistics to characterize the radius valley morphology (i.e., $C_{\rm valley}$, $A_{\rm valley}$, ${R}^{+}_{\rm valley}$, ${R}^{-}_{\rm valley}$, and $f_{\rm NP}$) as functions of $\rm [\alpha/Fe]$.
{\cdc In section \ref{sec.dis.mp}, we rescale the planet radii to eliminate the influence of stellar mass and planetary orbital period on the boundary of radius valley. The results with rescaling radii are plotted as open circles and shifted to the right to avoid overlapping.}
\label{DvalleyRENalpha}}
\end{figure}

{\jw 
Figure \ref{figRadiusValleyalpha} shows the period-radius diagram (upper panel) and radius distribution (bottom panel) of planets in different $\rm [\alpha/Fe]$ bins after parameter control.
As printed at the top, from left to right panels, the median values of $\rm \rm [\alpha/Fe]$ increases continuously while the kinematic age and $\rm [Fe/H]$ are nearly unchanged given their uncertainties. 
We see an apparent decay in  super-Earths accompanied by a rise in sub-Neptunes as $\rm [\alpha/Fe]$ increases.
This apparent trend is confirmed in the second panel of Figure \ref{DvalleyRENFeH}, which shows the asymmetry of the valley $\rm A_{valley}$ as a function of $\rm [\alpha/Fe]$; a linear decay model is prefered
than the constant model with a $\rm \Delta  AIC=13.75$ and a confidence level of 99.99\% from Monte Carlo simulation. The best-fit is 
  {\cdc  \begin{equation}
  A_{\rm valley}= -1.9^{+0.4}_{-0.1} \times {\rm [\alpha/Fe]} + 0.03^{+0.02}_{-0.02}.
    \label{eqA_alpha}
    \end{equation}}

Except for $\rm A_{valley}$, the other morphology metrics do not show a monotonic trend with $\rm [\alpha/Fe]$.
The $\rm C_{valley}$ seems to be a constant function of $\rm [\alpha/Fe]$.
For the other three metrics (i.e., ${R}^{+}_{\rm valley}$, ${R}^{-}_{\rm valley}$ and $f_{\rm NP}$ ), they all seem to follow a falling-rising V-shape pattern.
However, we caution that this pattern could be unphysical because there are only four data points (four bins) here, while a V-shape model needs at least four parameters.
Besides the small number of bins, the total range of $\rm [\alpha/Fe]$ (between -0.1 and 0.1) is also small and the differences in $\rm [\alpha/Fe]$ between adjacent bins are only $\sim$ 1 sigma (see the horizontal error bars in Figure \ref{figRadiusValleyalpha}).
All these limitations add difficulty to further reveal the physical connection between radius valley morphology and $\rm [\alpha/Fe]$.

{\cdc In all the above investigations, we eliminate the influence of the observational bias indirectly via constructing control samples.
In Appendix \ref{sec.Appen.DE}, we calculated the transit detection efficiency and found that the mean detection efficiency, or completeness, does not
change substantially for our controlling samples across all $TD/D$, age, $\rm [Fe/H]$ and $\rm [alpha/Fe]$ (as shown in Figure B1-B4).
Therefore, we conclude that completeness corrections are not likely to (significantly) impact the main results of our analysis.}
{\CDC Since previous studies from observational analyses \citep[e.g.,][]{2020AJ....160..108B,2021ApJ...911..117S} and theoretical models \citep[e.g.,][]{2020MNRAS.493..792G,2021MNRAS.503.1526R} all adopt the raw number ratio of super-Earth to sub-Neptunes, here we also adopt the raw number ratio to facilitate subsequent comparisons and discussions in section \ref{sec.dis}. }

}

\section{Discussions}
\label{sec.dis}
In this paper, {\chen based on the LAMOST-Gaia-Kepler catalog of PAST \uppercase\expandafter{\romannumeral2}\citep{2021AJ....162..100C}, {\jw we  study the radius valley in the Galactic context.
Specifically, we investigated the dependence of the radius valley morphology on Galactic component( $TD/D$), kinematic age and metallicity ($\rm [Fe/H]$ and $\rm [\alpha/Fe]$).}
The main results are summarized in Table \ref{tab:RadiusvalleyDependence}.
In this section, we will discuss our results from the following aspects. 
First, we {\chen discuss the variation of the radius valley boundary and its effect to our results (Sec. \ref{sec.dis.mp}).
Then, we compare our results to previous relevant works (Sec. \ref{sec.dis.comp}).
Finally, we discuss the implications of our results to planet formation and evolution (Sec. \ref{sec.dis.impl}).} 
}

\subsection{Boundaries of the Radius Valley}
\label{sec.dis.mp}
{In this paper, the radius valley is nominally and simply defined at $R^{\rm valley}_{\rm p0}=1.9 \pm 0.2 R_{\oplus}$.
In fact, the center of radius valley is found to be dependent on stellar mass  \citep{2019ApJ...874...91W,2020AJ....160..108B} and planetary orbital period  \citep{2018MNRAS.479.4786V}, which can be characterized as:
\begin{equation}
    R^{\rm valley}_{\rm p} = R^{\rm valley}_{\rm p0}
    \left(\frac{M_*}{M_\odot}\right)^h
    \left(\frac{P}{\rm 10 days}\right)^g,
\end{equation}
where $P$ and $M_*$ are planetary orbital period and stellar mass, and  the $h = 0.26^{+0.21}_{-0.16}$ \citep{2020AJ....160..108B} and $g = -0.09^{+0.02}_{-0.04}$ \citep{2018MNRAS.479.4786V} are the corresponding slopes, respectively.}

{\jw To take the above dependence into account, following \cite{2021arXiv210302127Z}, we re-scaled all planet radii as 
\begin{equation}
    \Tilde{R_{\rm p}} \equiv R_{\rm p} \left(\frac{M_*}{M_\odot}\right)^{-h}
    \left(\frac{P}{\rm 10 days}\right)^{-g},
\end{equation}
where $h$ and $g$ were set as 0.26 and -0.09 respectively.
With the re-scaled radii, then the boundaries (Sec. 2.1) used to classify different planet populations should be updated accordingly, i.e.,
}
\begin{enumerate}
    \item Valley planet (VP): $1.7 R_{\oplus} \le \Tilde{R_{\rm p}} \le 2.1 R_{\oplus}$;
    \item Super-Earth (SE): $R_{\rm p} \ge 1 R_{\oplus} {\rm \, and \,} \Tilde{R_{\rm p}}<1.7 R_{\oplus}$;
    \item Sub-Neptune (SN): $\Tilde{R_{\rm p}}>2.1 R_{\oplus} {\rm \, and \,} R_{\rm p} \le 3.5 R_{\oplus}$;
    \item Neptune-size planet (NP): $3.5 R_{\oplus}< R_{\rm p} \le 6 R_{\oplus}$.
\end{enumerate}
Note, here the boundaries of the Neptune-size planets are not affected by the changing in the valley boundaries.

{\jw With the above new criteria, we recalculated the five metrics of radius valley morphology, which are}
plotted as hollow grey circles in Figure \ref{DvalleyRENTDD}, \ref{DvalleyRENAge}, \ref{DvalleyRENFeH} and \ref{DvalleyRENalpha}, respectively.
As can be seen, the recalculated $C_{\rm valley}$, $A_{\rm valley}$, $f_{\rm NP}$, ${R}^{+}_{\rm valley}$ and ${R}^{-}_{\rm valley}$ (open circles) agree well with the our nominal results (solid points).
To quantitatively characterize the influence of stellar mass and planetary period on radius valley, we also fit those grey data points with the same procedure described in section \ref{sec.res}.
The results are all consistent with our nominal results within $1-\sigma$ uncertainty.

{\jw Based on the above analysis, we therefore conclude that our results (e.g., Figure \ref{DvalleyRENTDD}, \ref{DvalleyRENAge}, \ref{DvalleyRENFeH} and \ref{DvalleyRENalpha}) are not (significantly) affected by the dependence of radius valley on stellar mass and planetary orbital period.}
{\CDC Here we did not implement the rescaled radii from the beginning because the boundaries between sub-Earth and super-Earth, sub-Neptune and Neptune should not be scaled. Once taking the rescaled radii, the above boundaries should be different for different planets and cannot be easily plotted in Figure \ref{figRadiusValleyTDD}, \ref{figRadiusValleyAge}, \ref{figRadiusValleyFeH} and \ref{figRadiusValleyalpha}.}

\subsection{Comparisons to Other Studies}
\label{sec.dis.comp}

{\jw 
\subsubsection{studies of age dependence}
During writing this paper, we noted some recent works have also investigated the dependence of  planetary size distribution on stellar age.}
\cite{2020AJ....159..280B} used the isochrone age from the Gaia-Kepler stellar properties catalog and found that the number ratio of super-Earths to sub-Neptunes increases from $0.61 \pm 0.09$ for age $< 1$ Gyr to $1.00 \pm 0.10$ for age $> 1$ Gyr.
Shortly afterward, using the isochrone age from the CKS sample, \cite{2021ApJ...911..117S} shows that the number ratio of super-Earths to sub-Neptunes rise monotonically with age, from $0.76 \pm 0.08$ at $< 1$ Gyr to $0.94 \pm 0.07$ at $>1$ Gyr.
{\jw More recently, \cite{2021AJ....161..265D} reported that the radius valley appears emptier in younger systems ($< 2-3$ Gyr) but more filled in older systems.
However, such a trend is not seen in \cite{2020AJ....160..108B} nor in \cite{2021ApJ...911..117S}.
}

{\jw In this paper, we address the age dependence from a different angle of view by using the kinematic age based on the  LAMOST-Gaia-Kepler catalog \citep{2021AJ....162..100C}.
By defining a series of metrics, we  systematically investigated how the radius valley morphology evolves with age.
One of the metrics, $A_{\rm valley}$, measuring the asymmetry on the two sides of the radius valley, is defined as the logarithm of the number ratio of super-Earths to sub-Neptunes (Eq.\ref{eqREN}). 
We found $A_{\rm valley}\sim0$ for age $<1$ Gyr and it increases to $\sim0.2$ for age $>1$ Gyr (Figure \ref{DvalleyRENAge}), corresponding to an increase in the number ratio of super-Earths to sub-Neptunes from $\sim1.0$ to $1.5$.
The relative increase by $\sim50\%$ is a bit larger than that of \citet{2021ApJ...911..117S} but agrees well with that of \citet{2020AJ....160..108B}.
In addition, we note that the increase of the number ratio of super-Earths to sub-Neptunes seems mainly completed within the first $\sim 1-2$ Gyr, which is consistent with the result of \citet[][see their Figure 3]{2021ApJ...911..117S}.
The absolute value of the number ratio super-Earths to sub-Neptunes is systematically higher in this work.
Such a systematic difference may not be unexpected as neither the criteria to select the whole sample nor to define the super-Earths and sub-Neptunes is the same. 
For example, \citet{2020AJ....160..108B} focus on stars between 5700-7900 K and defines sub-Neptunes as planets between $1.8 \rm R_{\oplus}$-$3.5 \rm R_{\oplus}$.
While in this paper, we do not adopt any  stellar temperature cut, and thus our sample have relatively more cooler stars and smaller planets.
We also define sub-Neptunes in a smaller radius range between $2.1 \rm R_{\oplus}$-$3.5 \rm R_{\oplus}$.
Therefore, it is expected that the number ratio super-Earths to sub-Neptunes should be systematically higher in this work than in \citet{2020AJ....160..108B}.
}

{\jw
Like \citet{2020AJ....160..108B} and \citet{2021ApJ...911..117S}, we did not see the trend either, i.e, radius valley appears  emptier in younger systems than in older systems as claimed by \cite{2021AJ....161..265D}.
Instead, by defining another metric $\rm C_{Valley}$ to measure the contrast of the radius valley, we found an opposite trend, i.e., the valley becomes more striking with increasing age (Fig. \ref{DvalleyRENAge}).
We note that some key differences in the methodology could be the cause of the above disagreement. 
In this paper, in order to isolate the age effect, we have controlled other stellar parameters to have similar distributions across different age bins (Fig. \ref{figCDFMRCDPPAge}), while such a parameter control process has not been conducted in \cite{2021AJ....161..265D}.
As can be seen in the Figures 2 and 3 of \cite{2021AJ....161..265D}, stars in the youngest bin are mostly confined to large metallicity ($\rm [Fe/H]>0$), while stars in the oldest bin have much broader distributions in $\rm [Fe/H]$.
Given the significant dependence of radius valley on $\rm [Fe/H]$ as shown in Sec.\ref{sec.res.FeH} and also in \cite{2018MNRAS.480.2206O} (see more discussion below), 
we therefore suspect that the age trend observed in \cite{2021AJ....161..265D} could be affected by the effects of other stellar properties, e.g., $\rm [Fe/H]$.
}

{\jw \subsubsection{studies of metallicity dependence}
The $\rm [Fe/H]$ dependence of radius distribution of small planets ($\rm <6R_{\oplus})$ have been investigated by some previous studies.
With the LAMOST spectra, \citet{2018PNAS..115..266D} found that, {\cdc similar to hot Jupiter, hot planets} ($P<10 d$) with radii between 2-6$R_\oplus$ (dubbed as ''Hoptune") are mostly singles and  preferentially around stars with higher $\rm [Fe/H]$.
This trend was then confirmed by \citet{2018AJ....155...89P} with the Keck spectra, which further demonstrated that there is a greater diversity of planets around higher $\rm [Fe/H]$ stars.
With the CKS sample, \cite{2018MNRAS.480.2206O} studied the $\rm [Fe/H]$ dependence of radius distribution for planets around the radius valley.  
They found that the radius distribution of planets above the valley ($\gtrsim 1.8 R_\oplus$) is shifted to larger radii around higher $\rm [Fe/H]$ stars, leading to a wider radius valley around higher-$\rm [Fe/H]$ stars.
This can be clearly seen from their Figure 2, which also reveals that the ratio of super-Earths to sub-Neptunes is smaller for higher-$\rm [Fe/H]$ stars.  
All their findings are consistent with our results (Fig. \ref{DvalleyRENFeH}). 
Their results on the radius shift of planets above the valley agree with our results on the increase in $R^{+}_{\rm valley}$ and $f_{\rm Nep}$ with $\rm [Fe/H]$.
Their results on the wider valley and smaller ratio of super-Earths to sub-Neptunes for higher-$\rm [Fe/H]$ stars agree with our results on the increase in $\rm C_{Valley}$ and decrease in $A_{\rm valley}$ with $\rm [Fe/H]$, respectively.

{\chen However, with the CKS \uppercase\expandafter{\romannumeral7} sample \citep{2018AJ....156..264F}, recently \cite{2021AJ....162...69K} argued that the observed correlation between stellar metallicity and planetary radius could be explained by the} {\jw combination of another two correlations, i.e., the stellar mass-metallicity correlation and the stellar mass-planetary radius correlation.
Their result is inconsistent with ours, in which there is still significant correlation between stellar metallicity and planetary radius even after parameter control in stellar mass.
In order to understand the reasons for this discrepancy, we performed the following two exercises. 

\begin{enumerate}

\item  We downloaded the CKS \uppercase\expandafter{\romannumeral7} sample and conducted similar correlation analyses as in \cite{2021AJ....162...69K}. 
Figure \ref{figCKS7SPC} reproduces the correlations of $M_*-\rm  [Fe/H]$, $M_*-R_{\rm p}$ and $\rm [Fe/H]-R_{\rm p}$ as seen in Figure 1 of \cite{2021AJ....162...69K} quantitatively. 
Based on these correlations, \cite{2021AJ....162...69K} argues that the $M_*-\rm [Fe/H]$ and $M_*-R_{\rm p}$ correlations are intrinsic while the $\rm [Fe/H]-R_{\rm p}$ correlation is just a projection of the former two.
Nevertheless, here we find that the situation is more complicated.
As shown in Figure \ref{figCKS7SPC}, there are more correlations which were not considered in \cite{2021AJ....162...69K}, namely the stellar age is anti-correlated with stellar mass and metallicity, which is generally expected from the theory of star formation and evolution \citep[e.g.,][]{2014A&A...562A..71B,2021ApJ...909..115C}.
In fact, all these observed correlations are not quantitatively intrinsic because any of them is more or less affected by the others.
Therefore, these \emph{apparent} correlations should be taken with caution and they are not reliable for direct use and interpretation. 

\item Bearing this in mind, we then apply parameter control to the above CSK \uppercase\expandafter{\romannumeral7} sample.
Specifically, following the same procedure as in section 3.3, we divided the whole sample into four bins according to their $\rm [Fe/H]$ and used the Nearest Neighbour method to select the nearest neighbor in the space of the controlled parameters (i.e., stellar mass, radius, age and CDPP) from stars in the latter three bins for every star in the first bin. 
After such a parameter control process, stars in different $\rm [Fe/H]$ bins have similar distributions in mass, radius and age (as shown in Figure \ref{figCDFMRCDPPFeHCKS7}), and as expected, the mutual correlations among mass, radius and age become much weaker (as shown in Fig. \ref{figCKS7SPCAPCFeH}, all $p$ values are larger than 0.05). 
In contrast, the correlation between $\rm [Fe/H]$ and $\rm R_p$ still maintains as significant ($p$ value $=10^{-4}$) as before.
\end{enumerate}

To summarize the above two exercises, we learn that the reason for the discrepancy is not in the different samples but in the different methods used between \cite{2021AJ....162...69K} and this work.
The former reaches their conclusion by directly analyzing the \emph {apparent} correlations, which we argue could be unreliable.
Our work attempts to extract the intrinsic correlation by controlling other parameters to isolate the effect of $\rm [Fe/H]$.
}

As for $\rm [\alpha/Fe]$, early studies \citep[e.g.,][]{2002AJ....124.2224Z} found that giant planet host stars show a slight overabundance for some $\alpha$ elements.
Later, the overabundance of $\alpha-$elements is found to be most striking for iron poor regime \citep{2008MNRAS.388.1175H,2011ApJ...736...87K}, and it exhibits not only in stars hosting giant planets but also in stars hosting Neptune and super-Earth class planets \citep{2012A&A...543A..89A}. 
Recently, \cite{2018ApJS..237...38B} found there is no significant difference in $\rm Mg/Si$ ratio between super-Earth hosts and sub-Neptune hosts.
In this paper, with the LAMOST-Gaia-Kepler catalog of PAST \uppercase\expandafter{\romannumeral2}, we found that $\alpha-$elements play an important role in shaping the radius valley, leading to a decrease in the asymmetry ($\dot{A}_{\rm valley}<0$, Fig.\ref{DvalleyRENalpha}) i.e., smaller number ratio of super-Earths to sub-Neptunes for stars with higher $\rm [\alpha/Fe]$.
}

\begin{table*}[!t]
\centering
\renewcommand\arraystretch{1.5}
\caption{Dependence of radius valley morphology (i.e., $C_{\rm valley}$, $A_{\rm valley}$,  ${R}^{+}_{\rm valley}$, ${R}^{-}_{\rm valley}$, and $f_{\rm NP}$, section \ref{sec.methods.statistics}) on  $TD/D$ (Sec. \ref{sec.res.TDD}), kinematic age (Sec. \ref{sec.res.kineage}), $\rm [Fe/H]$ (Sec. \ref{sec.res.FeH}) and $\rm [\alpha/Fe]$ (Sec. \ref{sec.res.alpha}).}
{\footnotesize
\label{tab:RadiusvalleyDependence}
\begin{tabular}{l|c|c|c|c|c} \hline
                &  $C_{\rm valley}$ (Eq. \ref{eqDV}) & $A_{\rm valley}$ (Eq. \ref{eqREN}) & ${R}^{+}_{\rm valley}$ (Eq. \ref{eqRNSE}) & ${R}^{-}_{\rm valley}$ (Eq. \ref{eqRSE}) & $f_{\rm NP}$ (Eq. \ref{eqfNep}) \\ \hline
   $\bm{TD/D  \uparrow}$ (Fig. \ref{DvalleyRENTDD}) & 
   $--$ &
   $--$ &
   $\bm{\downarrow}$ & 
   $\bm{\rightsquigarrow}$ & 
   $--$  \\
   $\rm \bm{Age \uparrow}$ (Fig. \ref{DvalleyRENAge})& 
   $\bm{\uparrow}$ (Eq. \ref{eqC_Age}) & 
   $\bm{\uparrow}$ (Eq. \ref{eqA_Age})& 
   $\bm{\downarrow}$ (Eq. \ref{eqR+_Age})&
   $\bm{\rightsquigarrow}$ (Eq. \ref{eqR-_Age})&
   $\bm{\downarrow}$ (Eq. \ref{eqfNP_Age}) \\
   $\rm \bm{[Fe/H] \downarrow}$ (Fig. \ref{DvalleyRENFeH}) & $\bm{\downarrow}$ (Eq. \ref{eqC_Fe})& 
   $\bm{\uparrow}$ (Eq. \ref{eqA_Fe})& 
   $\bm{\downarrow}$ (Eq. \ref{eqR+_Fe}) &
   $\bm{\rightsquigarrow}$ (Eq. \ref{eqR-_Fe}) & 
   $\bm{\downarrow}$ (Eq. \ref{eqfNp_Fe}) \\
   $\rm \bm{[\alpha/Fe]} \uparrow$ (Fig. \ref{DvalleyRENalpha}) & $\bm{\rightsquigarrow}$  & $\bm{\downarrow}$ (Eq. \ref{eqA_alpha})& 
   $--$ & $--$ & 
   $--$  \\ \hline
\end{tabular}}
\flushleft{\cdc 
$\uparrow$: increasing; $\downarrow$: decreasing; $\rightsquigarrow$: broadly unchanged; 
$--$: unclear.
} 
\end{table*}

\subsection{Implications to Planet Formation and Evolution}
\label{sec.dis.impl}

\subsubsection{radius valley emerged before Gyr and evolving beyond Gyr}
\label{sec.dis.impl.age}
{\jw 
In this paper (section \ref{sec.res.kineage}), we investigate the temporal evolution of radial valley and find both the contrast ($C_{\rm valley}$) and the asymmetry ($\rm A_{valley}$, or the ratio of super-Earths to sub-Neptunes equivalently) of the valley increase with age (the top two panels of Figure \ref{DvalleyRENAge}).
Qualitatively, our results suggest that the radius valley is evolving on giga years timescales, which generally supports the evolutionary models driven by photo-evaporation \citep[e,g.,][]{2013ApJ...775..105O,2014ApJ...795...65J, 2016ApJ...818....4L}  and/or cooling luminosity of planetary core \citep{2016ApJ...825...29G,2018MNRAS.476..759G,2019MNRAS.487...24G,2020MNRAS.493..792G}. 
Quantitatively, as we will discuss below, the radius valley morphology as a function of time (Eqn. 7), may suggest that both the phtoto-evaporation and core-powered mass-loss mechanisms should be at play, providing constraints to quantify their roles in planetary evolution. 


As shown in Figure \ref{figRadiusValleyAge} and Figure \ref{DvalleyRENAge}, the radius valley had already emerged before 0.5 Gyr. 
Afterwards, it continues evolving beyond giga years, with the number ratio of super-Earths to sub-Neptunes increasing by $\sim50\%$ from 0.45 Gyr to 4.58 Gyr. 
Neither the photo-evaporation mechanism alone or the core-powered mechanism alone can fully explain the above observational results. 
On one hand, although the photo-evaporation mechanism can readily form a radius gap well before 0.5 Gyr, the increase in the ratio of super-Earths to sub-Neptunes is somewhat too small \citep[only $\sim20\%$ from $0.77 \pm 0.08$ at $t<1$ Gyr to $0.95 \pm 0.08$ at $t>1$ Gyr as shown in the figure 16 of][]{2021MNRAS.503.1526R} because the photo-evaporation operates mostly within the first few 100 Myr \citep[although see][]{2021MNRAS.501L..28K}.
On the other hand, although core-powered mechanism operates on a much longer timescale of a few Gyr and thus causes a sufficient increase in the ratio of super-Earths to sub-Neptunes over giga years (from 0.06 to 0.52 at ages of 0.5 and 3 Gyr by a factor of $\sim 9$ \citep[see fig.10 of][]{2020MNRAS.493..792G}, the radius valley under this mechanism emerges somewhat too late \citep[after $\sim1$ Gyr as shown in][]{2020MNRAS.493..792G}.
Nevertheless, a hybrid model which combines both the photo-evaporation and core-powered mechanisms seems promising to fully explain our results. 
The photo-evaporation and core-powered mechanisms could be complementary to each other, namely, the former dominates the early formation and evolution of the valley while the latter further strengthens the long-term evolution of the valley. 
Future detailed studies will test and further constrain the hybrid model.
}

\subsubsection{constraints on planetary thermodynamic evolution}
\label{sec.dis.impl.thermal}
{\jw Planets born with significant gaseous envelope are expected to significantly shrink in size due to long term thermodynamic evolution, i.e., planet cooling and contraction \citep{2011ApJ...729...32F,2019A&A...623A..85L}.
The evidence of such a thermodynamic evolution is clearly shown in our result in Figure 6.
As can be seen, both the fraction of Neptune size planets ($f_{\rm NP}$) and the average radius of planet above the radius valley ($R^+_{\rm valley}$) continuously decrease with age.
This result also implies that the bulk of planets above the radius valley ($\rm R_p>2.1R_\oplus$) should be made up with significant gaseous envelope (e.g., H/He atmosphere).
In contrast, planets below the radius valley ($\rm R_p<1.7R_\oplus$) do not show significant change in their sizes over Gyr, which may imply that they are made of mostly bare cores \citep{2014ApJ...783L...6W, 2015ApJ...801...41R}, in line with the expectation of the {\cdc atmosphere escape models \citep{2017ApJ...847...29O,2018ApJ...853..163J,2018MNRAS.476..759G}}.

Furthermore, we find that $R^{+}_{\rm valley}$ decreases with age by a slope of ${\rm d}(\log R)/{\rm d}(\log t)=-0.08^{+0.01}_{-0.01}$, which quantitatively places an observational constraint on the thermodynamic evolution of sub-Neptunes and Neptunes.
For comparison,  \citet{2020MNRAS.493..792G} found a slope of ${\rm d}(\log R)/{\rm d}(\log t) = \sim -0.1$ from their theoretical model by including the effect of core-powered atmosphere mass loss.
Nevertheless, the thermodynamic evolution depend not only on the core cooling but also stellar irradiation, e.g., photo-evaporation \citep{2021MNRAS.503.1526R}, planetary composition \citep{2016ApJ...831..180C} and etc.
{\cdc We also explore the average radii of sub-Neptunes ($2.1-3.5 \ R_\oplus$) and Neptunes ($3.5-6 \ R_\oplus$) separately and found they both decrease with increasing age continuously, suggesting that the envelope of sub-Neptunes and Neptunes all shrink with age.}
Future studies, by comparing our observational results to a comprehensive thermodynamic evolution model, could potentially reveal more insights on the compositions of sub-Neptunes and the roles of different mass-loss mechanisms.   
}


\subsubsection{effects of metallicity on planetary formation and evolution }
\label{sec.dis.impl.elements}
In section \ref{sec.res.FeH}, we have explored the dependence of radial distributions on $\rm [Fe/H]$ (Fig. 7-8). 
Our results show that the ratio of super-Earth to sub-Neptunes decreases with the increase of $\rm [Fe/H]$.
We also find the average radii of {\jw planets above the radius valley} is larger around metal-richer stars and thus forms a wider radius valley.
The number fraction of {\jw Neptune-sized planets} grows with increasing $\rm [Fe/H]$.

{\jw Qualitatively, the above results could} be explained by the following two scenarios.
First, from the view of planetary formation, planets around metal-rich stars {\jw are expected to} have more massive cores and accrete more gas, forming larger sub-Neptunes and/or Neptune-size planets  \citep[e.g.,][]{2012A&A...541A..97M,2018MNRAS.480.2206O,2020arXiv200705562E}. 
{\jw Second, from the view planetary thermodynamic evolution, planets are expected to cool and contract slower with higher opacity of the envelope. 
Assuming that the opacity is correlated to the metallicity of the protoplanetary disk as well as the $\rm [Fe/H]$ of host star, one would expect sub-Neptunes and Neptune-size planets} are more frequent and larger around stars with higher metallicity at the same age \citep[e.g.,][]{2020MNRAS.493..792G}.

Quantitatively, {\jw we find that the average radii of sub-Neptunes increases with $\rm [Fe/H]$ by a slope of ${\rm d} \log {R}^{+}_{\rm valley}/{\rm d} [\rm Fe/H] = 0.094^{+0.012}_{-0.011}$, which is well consistent with the planetary cooling simulations by \cite{2020MNRAS.493..792G} who found a slope of ${\rm d} \log {R}/{\rm d} [\rm Fe/H] \sim 0.1$.}
{\jw
Thus, in term of the slope, our results provide evidence to support the second scenario.
Nevertheless, this does not mean that the first scenario is ruled out.
In fact, the second scenario alone may not fully explain all the observations.
For instance, we note that the observed $R^{+}_{\rm valley}$ is symmetrically larger than the simulation results of \citet[][see also in their Fig. 6]{2020MNRAS.493..792G}}.
One possible explanation could be that \cite{2020MNRAS.493..792G} did not consider the effect of $\rm [Fe/H]$ on the core/gas accretions and the enlargement of planetary radius (potential evidence for the first scenario) during the process of planetary formation.


As for $\rm [\alpha/Fe]$ {\jw (section \ref{sec.res.alpha})}, we have found the number ratio of super-Earths to sub-Neptunes decreases significantly with increasing $\rm [\alpha/Fe]$.
That is to say that sub-Neptunes are preferentially around $\alpha-$rich stars. 
As $\alpha-$element abundances {\jw provided} by LAMOST are the average abundances of Mg, Si, Ca, and Ti elements, one possible explanation is that planets around $\alpha-$ rich stars could be born in disks with more silicates and thus have {\jw larger rocky cores} to accrete more gas, forming more sub-Neptunes than super-Earths.
Our result suggests that metal elements other than Fe (e.g., Mg, Si, Ca, and Ti) also {\jw play important roles in} the formation of sub-Neptunes.
On this basis, we predict that sub-Neptunes around {\jw $\alpha-$richer} stars with similar $\rm [Fe/H]$ will have {\jw relatively lower densities}.



\section{Summary}
\label{sec.sum}
Using the LAMOST-Gaia-Kepler kinematic catalogs provided by PAST \uppercase\expandafter{\romannumeral2} \citep{2021AJ....162..100C}, we explored the radius valley morphology of small planets (1-6 $\rm R_{\oplus}$) in the Galactic context.
Specifically, we {\jw define a set of metrics to quantify the radius valley morphology (Sec. \ref{sec.methods.statistics}) and investigate whether and how they are correlated with the properties of host stars, i.e., the relative membership probability between thick and thin disk stars (TD/D), the kinematic age and metallicity, i.e., $\rm [Fe/H]$ and $\rm [\alpha/Fe]$ (Sec. \ref{sec.res}).}
{\jw
We did not find any significant correlation between the radius valley and TD/D (Sec. \ref{sec.res.TDD}), which is not unexpected because TD/D itself strongly depends on stellar age and metallicity, and these properties could affect the radius valley in different ways.
After applying parameter control to isolate the effects of stellar age and metallicity, we then find a number of correlations, which are summarized as the follows and also in Table \ref{tab:RadiusvalleyDependence}.
}
{\jw 
\begin{itemize}
   \item
    For the dependence on age (Fig. \ref{figRadiusValleyAge}-\ref{DvalleyRENAge}), we found that the radius gap emerged before 1 Gyr and continued evolving on Gyr timescales.
    Specifically, both the contrast and the asymmetry (or the number ratio of super-Earths and sub-Neptunes equivalently) of the radial valley increase with age (Eqn. \ref{eqC_Age}-\ref{eqA_Age}).
   The fraction of Neptune-sized planets and the average radii of planets above the radius gap ($2.1-6 R\oplus$) are found to decrease with age (Eqn \ref{eqR+_Age} and \ref{eqfNP_Age}), while the average radii of planets below the radius valley ($<1.7R\oplus$) shows no significant dependence on age (Eqn. \ref{eqR-_Age}).
 
   \item For the dependence on $\rm [Fe/H]$, the contrast of the radius valley increases while the asymmetry (or the number ratio of super-Earths and sub-Neptunes equivalently) decreases with $\rm [Fe/H]$ (Eqn. \ref{eqC_Fe}-\ref{eqA_Fe}).
   The fraction of Neptune-sized planets and the average radii of planets above the radius gap ($2.1-6 R\oplus$) are found to increase with $\rm [Fe/H]$ (Eqn. \ref{eqR+_Fe} and \ref{eqfNp_Fe}), while the average radii of planets below the radius valley ($<1.7R\oplus$) are broadly unchanged (Eqn. \ref{eqR-_Fe}).
   
   \item For the dependence on $\alpha$ elements,
    we found the the asymmetry of the valley (or the number ratio of super-Earths and sub-Neptunes equivalently) decreases with $\rm [\alpha/Fe]$ (Eqn. \ref{eqA_Age}). 
    Apart from this, we did not found any significant correlation between $\rm [\alpha/Fe]$ and other metrics of the valley morphology, probably because the uncertainty of $\rm [\alpha/Fe]$ is relatively large compared to the range of $\rm [\alpha/Fe]$ in our sample.  
   
\end{itemize}
}

Our results on the long term temporal evolution of the radius valley support the evolutionary models driven by atmospheric mass loss due to photo-evaporation \citep{2013ApJ...775..105O,2014ApJ...795...65J} and cooling planetary cores \citep{2016ApJ...825...29G,2018MNRAS.476..759G,2019MNRAS.487...24G,2020MNRAS.493..792G}, and suggest that the radial valley is a result of planetary atmospheric loss over Gyr timescale, which is primarily formed by the photoevaporation at early stage and further strengthen by the combined effect of photoevaporation and core-powered mass loss (Sec. \ref{sec.dis.impl.age}). 
The dependence of radius distribution on age also provides evidences and constraints on planetary thermodynamic evolution, as well as insights on planetary compositions (Sec. \ref{sec.dis.impl.thermal}).
Planets above the radius valley ($2.1 R_\oplus< R_{\rm p}<6 R_\oplus$) should be made with significant H/He envelope, while those below the radius valley ($1 R_\oplus< R_{\rm p}<1.7 R_\oplus$) are likely to be bare rocky cores.  
The dependence of radius distribution on metallicity suggests that not only Fe but also other metal elements (Mg, Si, Ca, Ti and etc)  play important roles in the formation and evolution of super-Earths and sub-Neptunes (Sec. \ref{sec.dis.impl.elements}). 

Our work emphasizes the importance of parameter control in studying the links between various planetary properties and stellar characteristics. 
Future studies, both from observational analysis (e.g., very young stars with age less than 100 Myr) and numerical simulations of various theoretical models will test our results and further provide more insights and constraints on the formation, evolution and compositions of exoplanets.


\section*{Acknowledgements}
We thank Wei Zhu, Yanqin Wu, Fei Dai, Mao-Sheng Xiang and James E. Owen for helpful discussions and suggestions.
This work is supported by the National Key R\&D Program of China (No. 2019YFA0405100) and the National Natural Science Foundation of China (NSFC; grant No. 11933001, 11973028,  11803012, 11673011,  11903005, 12003027). J.-W.X. also acknowledges the support from the National Youth Talent Support Program and the Distinguish Youth Foundation of Jiangsu Scientific Committee (BK20190005).
D.-C.C. also acknowledges the Cultivation project for LAMOST Scientific Payoff and Research Achievement of CAMS-CAS.

This work has included data from Guoshoujing Telescope
(the Large Sky AreaMulti-Object Fiber Spectroscopic Telescope LAMOST), which is a National Major Scientific Project built by the Chinese Academy of Sciences. Funding for the project has been provided by the National Development and Reform Commission.
This work presents results from the European Space Agency (ESA) space mission Gaia. Gaia data are being processed by the Gaia Data Processing and Analysis Consortium (DPAC). Funding for the DPAC is provided by national institutions, in particular the institutions participating in the Gaia MultiLateral Agreement (MLA). 
We acknowledge the NASA Exoplanet Science Institute at IPAC, which is operated by the California Institute of Technology under contract with the National Aeronautics and Space Administration.

\newpage
\appendix
\renewcommand\thefigure{\Alph{section}\arabic{figure}}
\setcounter{figure}{0} 
\section{Construction and validation of the control bins}
\begin{figure*}[!h]
\centering
\includegraphics[width=0.9\textwidth]{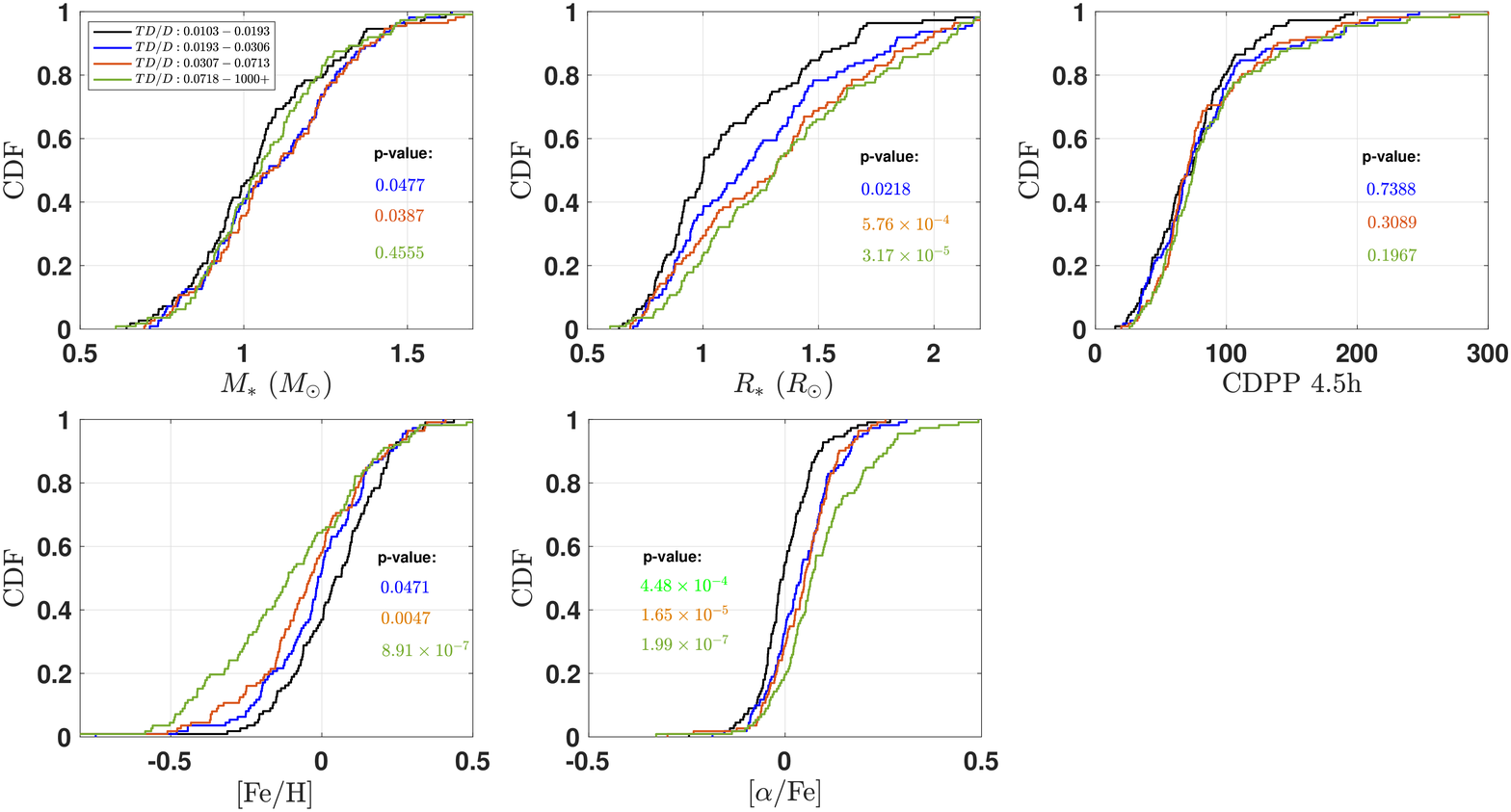}
\caption{\CDC The cumulative distributions of stellar mass $M_*$, radius $R_*$, CDPP 4.5h, $\rm [Fe/H]$ and $\rm [\alpha/Fe]$ for the four bins with \textbf{different relative probability between thick disk to thin disk $TD/D$ (Sec. \ref{sec.res.TDD})} before parameter controlling. 
Here we directed divide our sample into four sample with approxiamtely equal size ($\sim 111$) according to $TD/D$.
In each panel, the $p$ denotes the $p-$value of the two sample KS test for the distributions of neighboring star belonging to the latter three bins comparing to the first bins. 
\label{figCDFMRCDPPTDDbeforecontrol}}
\end{figure*}

The radius valley has been found to be correlated to multiple stellar properties (e.g., age, mass, $\rm [Fe/H]$).
These stellar properties are not independent but found to be interrelated.
For example, star with larger $TD/D$ (kinematic age) are found to be poorer in $\rm [Fe/H]$ \citep[e.g.,][]{2014A&A...562A..71B,2021ApJ...909..115C}.
The observation biases, caused by the detection efficiency, also need to be considered.
For example, the stellar activity and noise are anti-correlated to $TD/D$ and age \citep[Figure 5 and 6 of PAST \uppercase\expandafter{\romannumeral2};][] {2021AJ....162..100C}) and may affect the detection ability of planets.
To control the stellar detection ability and noise level,  here we adopt the Kepler DR25 Combined Differential Photometric Precision (CDPP) and the time scale is chosen as 4.5 hours as the calculated transit duration is about 4.7 hours if taking the median values of stellar mass, radius and planetary period in our sample.
{\CDC For example, we divided the sample into four equal-sized bins ($\sim$ 111 planets) according to $TD/D$ and compare their cumulative distributions of stellar parameters. As shown in the Figure \ref{figCDFMRCDPPTDDbeforecontrol}, with the increase of $TD/D$, both the stellar radius and CDPP (i.e., noise levels) grow, thus reducing planet detection efficiencies and causing bias toward larger planets for stars with larger $TD/D$. This is expected as stars with larger $TD/D$ are more likely to be in the thick disk and thus older and more distant. 
Therefore, in order to isolate the influence of a single property, one need to eliminate the influences of other properties and detection efficiency.}

To do this, we divide the whole sample in section \ref{sec.methods.sample} into four subsamples according to the studied property (i.e., $TD/D$, Age, $\rm [Fe/H]$ and $\rm [\alpha/Fe]$).
We take 10\% (44) of stars with the minimum $TD/D$, youngest kinematic age, highest $\rm [Fe/H]$ and lowest $\rm [\alpha/Fe]$ as the first bins for the section \ref{sec.res.TDD}, \ref{sec.res.kineage}, \ref{sec.res.FeH} and \ref{sec.res.alpha}, respectively. 
The subsequent samples are divided into three bins of equal size (134). 
Then we construct control samples by adopting the NearestNeighbors function in scikit-learn \citep{10.5555/1953048.2078195} to select the nearest neighbor in controlled properties for every star belonging to the first bin from stars belonging to latter three bins. 
In the case that a star is selected as the nearest neighbors of multiple times, we exclude duplicate stars.

\begin{figure*}[!h]
\centering
\includegraphics[width=0.9\textwidth]{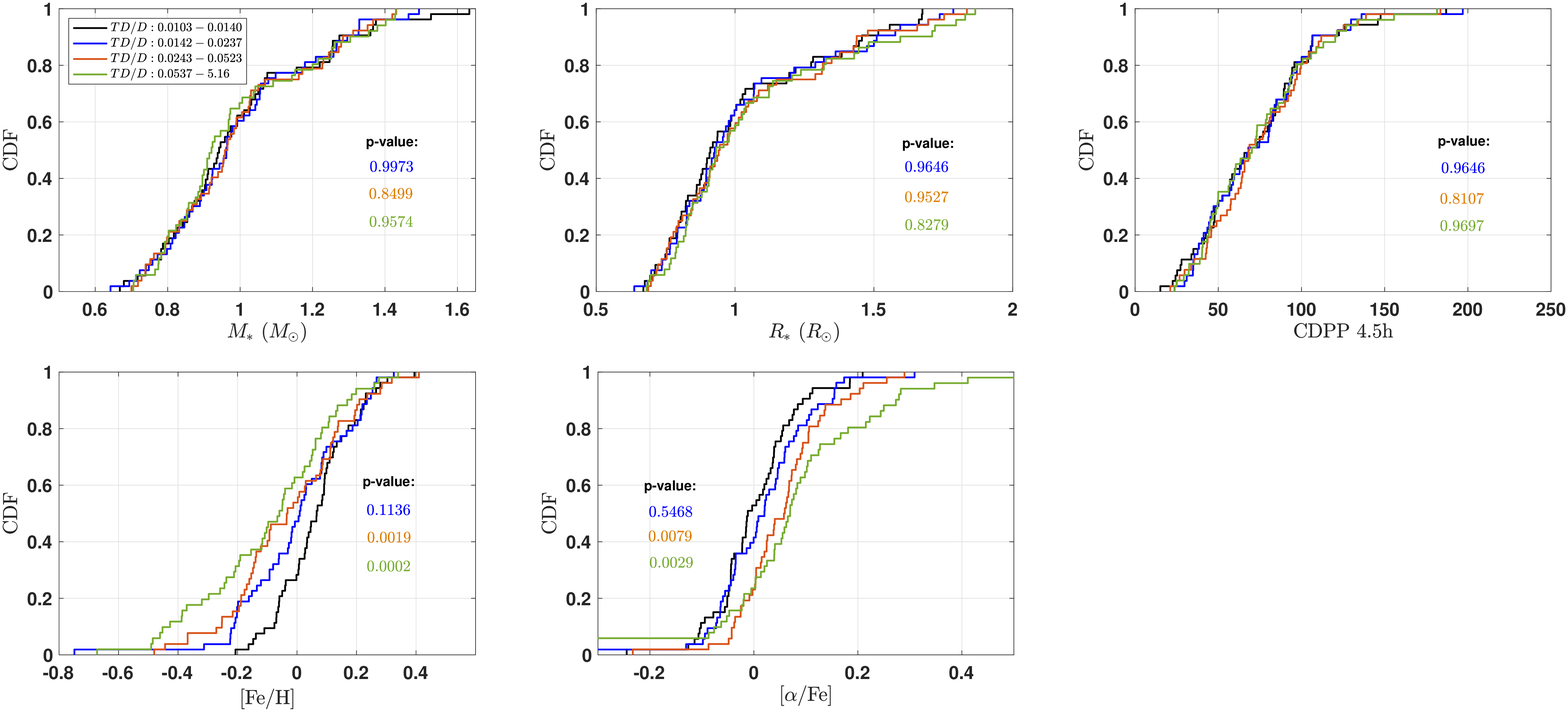}
\caption{The cumulative distributions of stellar mass $M_*$, radius $R_*$, CDPP 4.5h, $\rm [Fe/H]$, and $\rm [\alpha/Fe]$  for the four bins with \textbf{different relative probability between thick disk to thin disk $TD/D$ (Sec. \ref{sec.res.TDD})}. 
In each panel, the $p$ denotes the $p-$value of the two sample KS test for the distributions of neighboring star belonging to the latter three bins comparing to the first bins. 
Note, only stellar mass $M_*$, radius $R_*$, CDPP and Kepler magnitude have been controlled here.
We did not control $\rm [Fe/H]$ and $\rm [\alpha/Fe]$ when exploring the dependence on TD/D here, because the differences in stellar metallicity between thin and thick disk stars are essential and inevitable.
\label{figCDFMRCDPPTDD}}
\end{figure*}

\begin{figure*}[!h]
\centering
\includegraphics[width=0.9\textwidth]{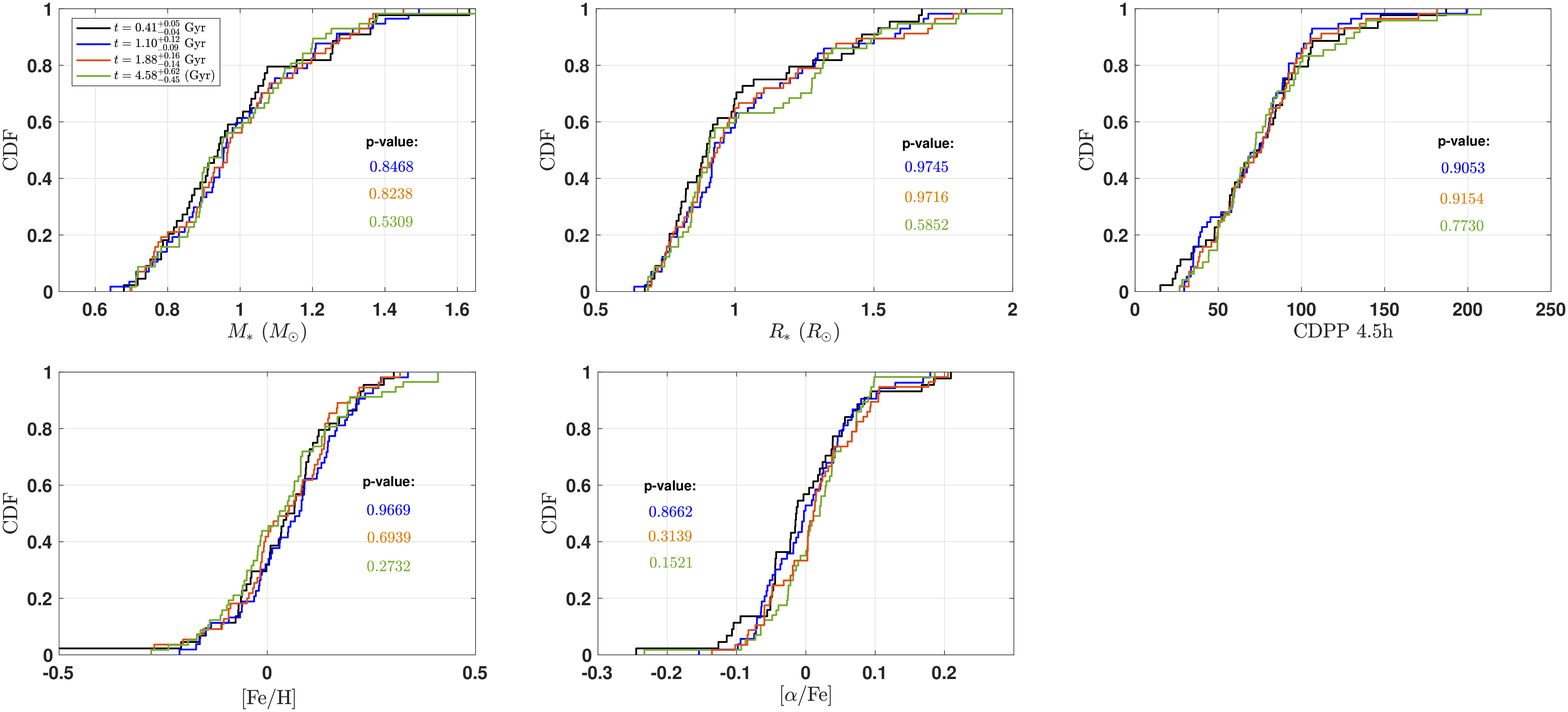}
\caption{Similar to Figure \ref{figCDFMRCDPPTDD}, but here shows the cumulative distributions of stellar mass $M_*$, radius $R_*$, CDPP 4.5h, $\rm [Fe/H]$, and $\rm [\alpha/Fe]$  for the four bins with \textbf{different kinematic age (Sec. \ref{sec.res.kineage})}. 
Note, here stellar mass $M_*$, radius $R_*$, CDPP, $\rm [Fe/H]$, $[\alpha/Fe]$ and Kepler magnitude have all been controlled in order to isolate the effect of age.
\label{figCDFMRCDPPAge}}
\end{figure*}

\begin{figure*}[!h]
\centering
\includegraphics[width=0.9\textwidth]{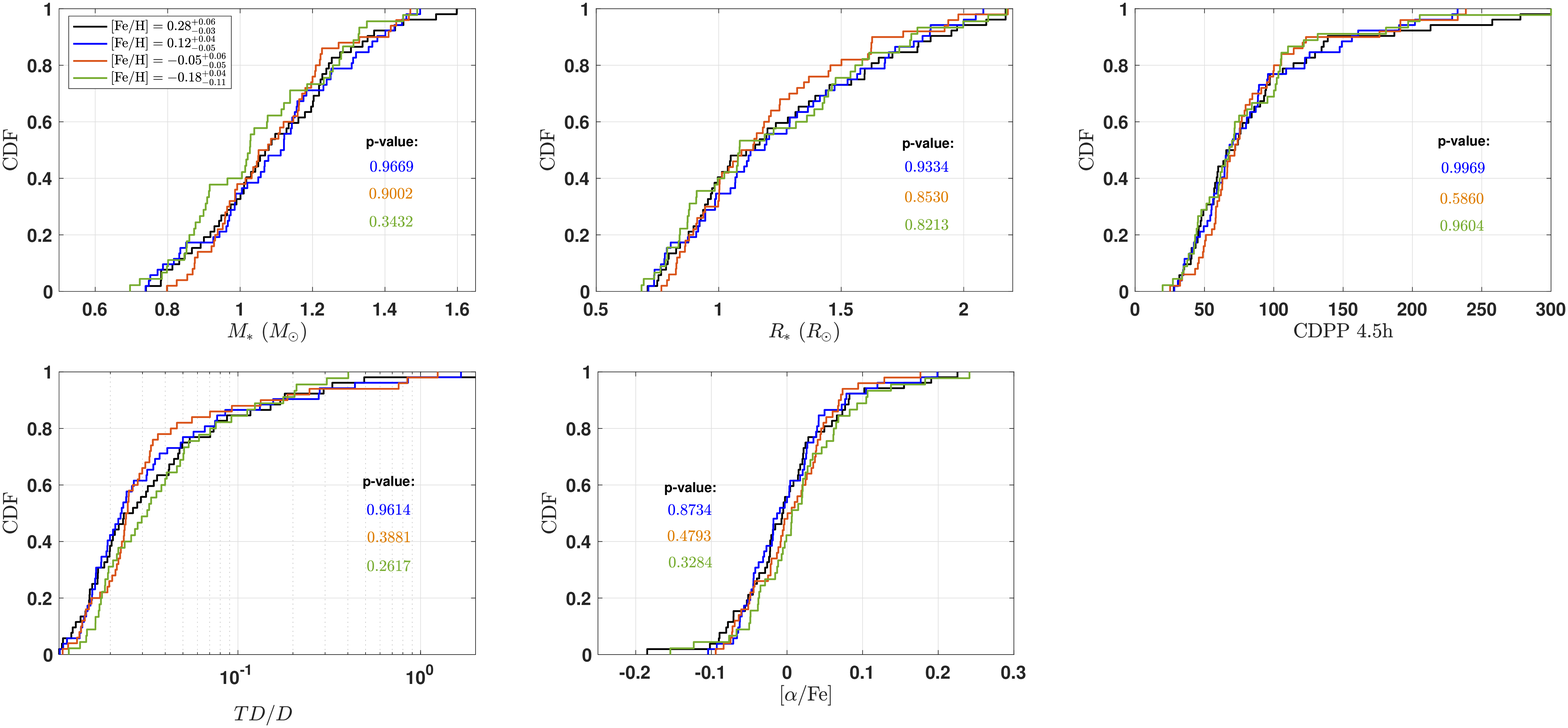}
\caption{Similar to Figure \ref{figCDFMRCDPPTDD}, but here shows the cumulative distributions of stellar mass $M_*$, radius $R_*$, CDPP 4.5h, $\rm TD/D$, and $\rm [\alpha/Fe]$ for the four bins with \textbf{different $\rm [Fe/H]$ (Sec. \ref{sec.res.FeH})}. 
Note, here stellar mass $M_*$, radius $R_*$, CDPP, $\rm TD/D$, $[\alpha/Fe]$ and Kepler magnitude have all been controlled in order to isolate the effect of $\rm [Fe/H]$. 
\label{figCDFMRCDPPFeH}}
\end{figure*}

\begin{figure*}[!h]
\centering
\includegraphics[width=0.9\textwidth]{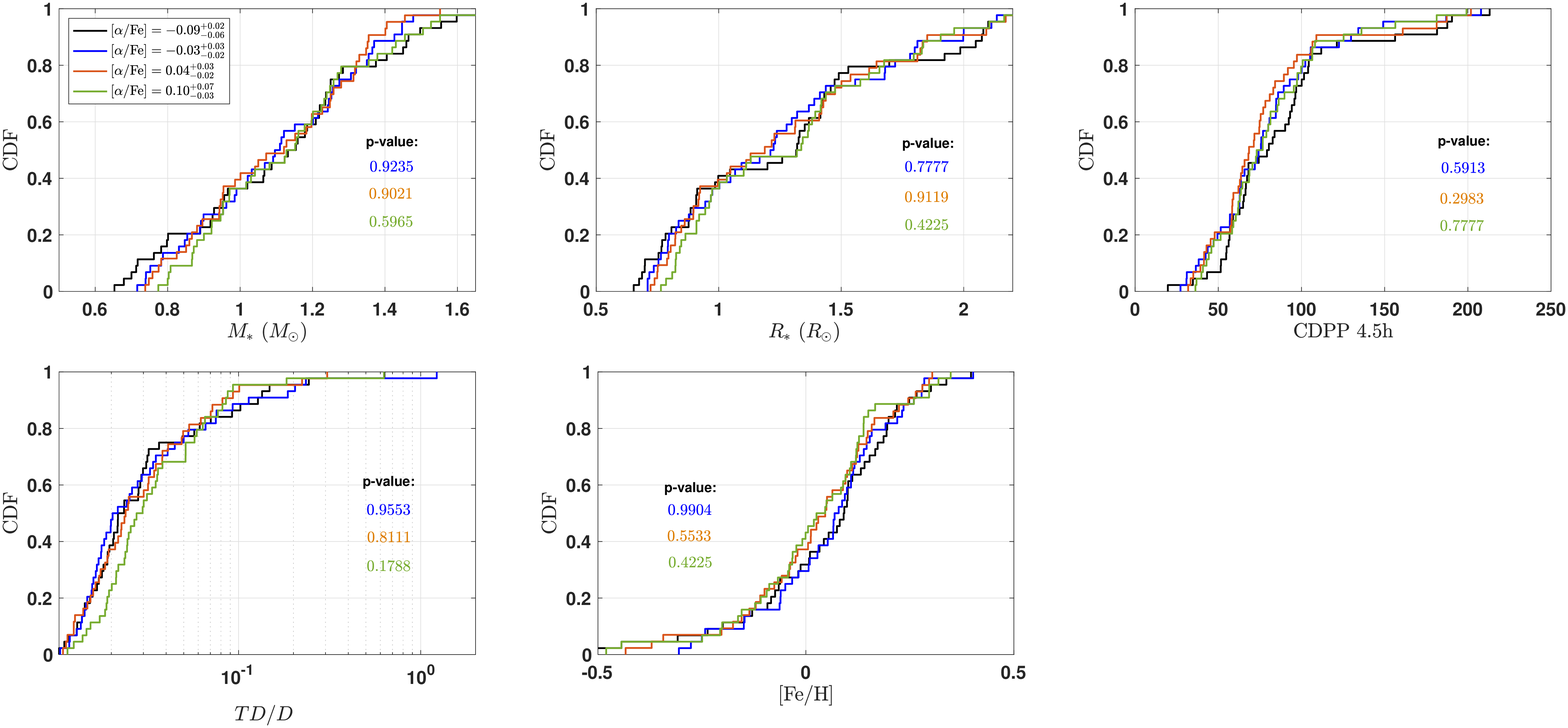}
\caption{Similar to Figure \ref{figCDFMRCDPPTDD}, but here shows the cumulative distributions of stellar mass $M_*$, radius $R_*$, CDPP 4.5h, $\rm TD/D$, and $\rm [Fe/H]$ for the four bins with \textbf{different $\rm [\alpha/Fe]$ (Sec. \ref{sec.res.alpha})}. 
Note, here stellar mass $M_*$, radius $R_*$, CDPP, $\rm TD/D$ and $\rm [Fe/H]$  have all been controlled in order to isolate the effect of $[\alpha/Fe]$. 
\label{figCDFMRCDPPalpha}}
\end{figure*}

To examine whether the control samples have similar distributions in other stellar properties, we make Kolmogorov–Smirnov (KS) tests and check the resulted $p-$values.
As can be seen in the Figure \ref{figCDFMRCDPPTDD}, \ref{figCDFMRCDPPAge}, \ref{figCDFMRCDPPFeH} and \ref{figCDFMRCDPPalpha}, all the KS test $p-$values of controlled parameters are larger than 0.15, demonstrating that these parameters have been well controlled and thus do not differ significantly in their distributions.
Note that in Figure \ref{figCDFMRCDPPTDD}, we did not control $\rm [Fe/H]$ and $\rm [\alpha/Fe]$ when exploring the dependence on $TD/D$, because the differences in stellar metallicity between thin and thick disk stars are essential and inevitable.

\begin{figure*}[!h]
\centering
\includegraphics[width=0.9\textwidth]{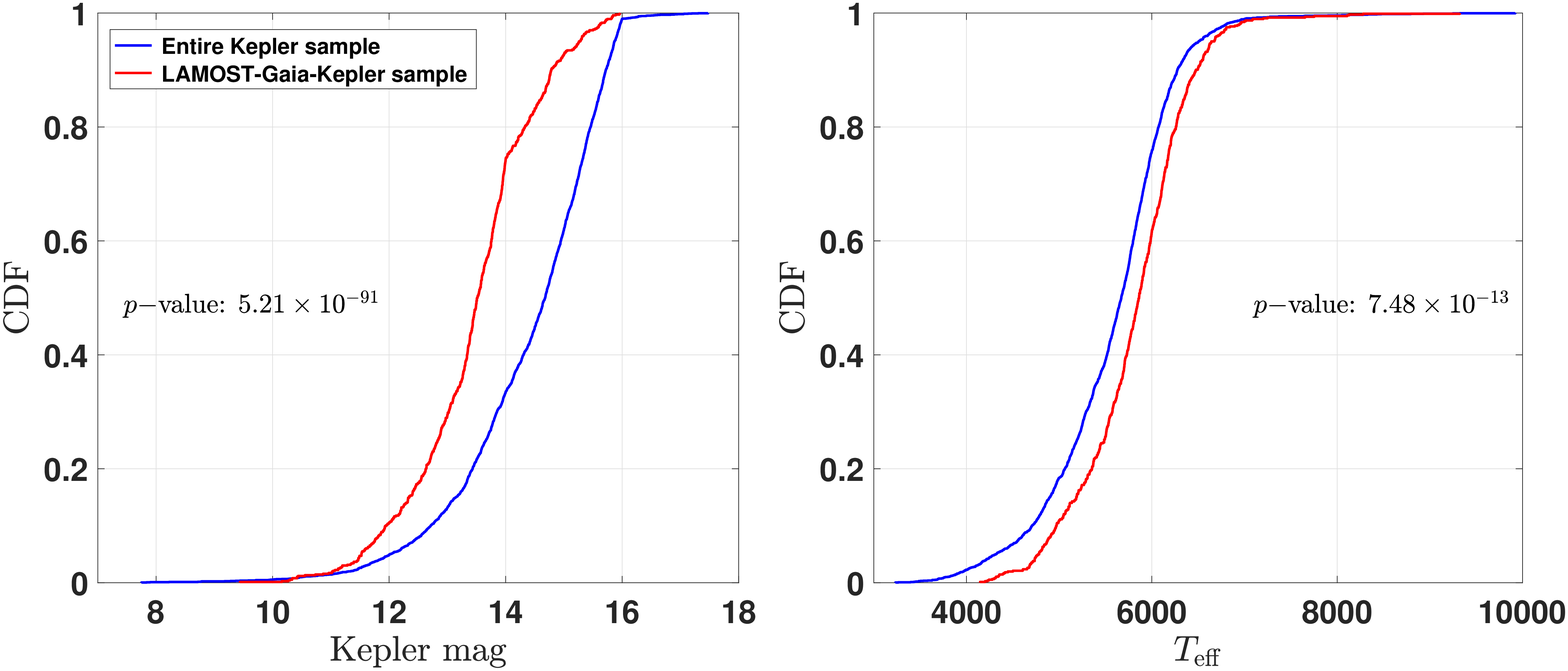}
\caption{\CDC The cumulative distributions of Kepler magnitudes and Teff for the planet host stars in the entire Kepler sample (blue) and LAMOST-Gaia-Kepler sample (red).
\label{figmagTeffLGKvsK}}
\end{figure*}

{\CDC It is also necessary to test observational biases of the LAMOST-Gaia-Kepler sample from PAST \uppercase\expandafter{\romannumeral2} \citep{2021AJ....162..100C}.
In Figure \ref{figmagTeffLGKvsK}, we
compare the distributions of the apparent magnitude and effective temperature of the planet host stars in the LAMOST-Gaia-Kepler sample and the entire Kepler sample.
As can be seen, planet hosts in the LAMOST-Gaia-Kepler sample have relatively smaller apparent Kepler magnitude and temperature. This is not unexpected because we present the LAMOST-Gaia-Kepler sample by cross-matching Kepler target stars with Gaia and LAMOST and LAMOST observed brighter stars preferentially as faint stars are harder to be accurately characterized.
Previous studies have shown that the radius valley morphology changes with magnitude \citep[e.g., Kepler magnitude $\lesssim 14$ vs. $\gtrsim 14$;][]{2017AJ....154..109F} and temperature \citep[a proxy for mass; e.g.,][]{2020AJ....160..108B}. Therefore, it is necessary to eliminate the influences of magnitudes and mass.
We have already controlled the stellar mass (corresponding to temperature) and radius (which combined corresponding to magnitudes), which could eliminate their influences. 
For further verification and visual display, we also check the cumulative distributions of Kepler magnitude and temperature. After parameter controlling, different bins of different $TD/D$ (Age, [Fe/H] and [alpha/Fe]) do not differ significantly in the distribution of Kepler magnitude and temperature.
}

\clearpage
\clearpage
\setcounter{figure}{0}
\section{ \cdc Detection Efficiencies of Different Bins of Various Properties}
\label{sec.Appen.DE}

\begin{figure*}[!h]
\centering
\includegraphics[width=0.6\textwidth]{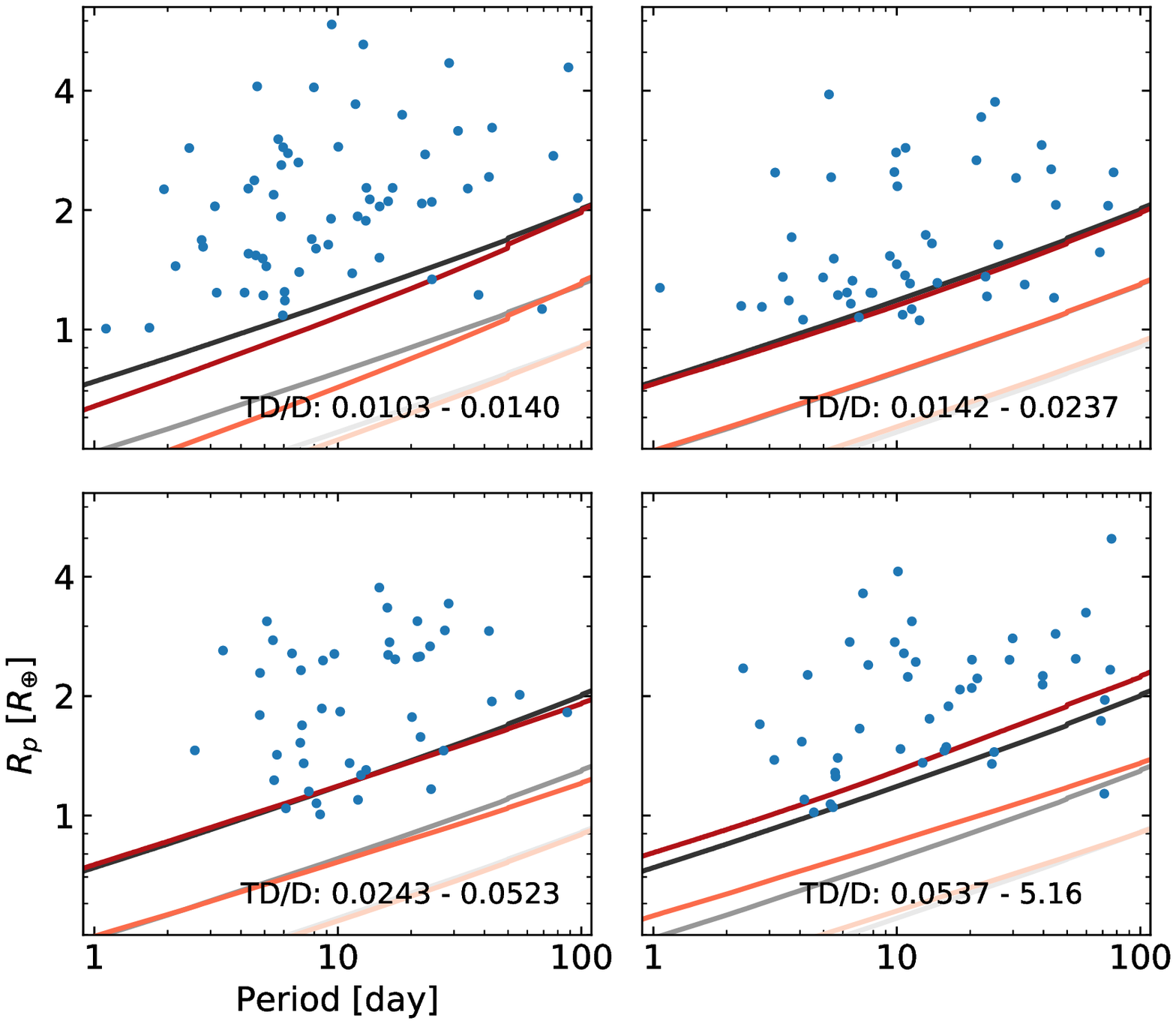}
\caption{Orbital periods vs. radii for the planets in the four bins of different TD/D after parameter controlling. In each panel, we plot their average detection efficiencies (10\%, 50\%, and 90\%) for stars in each bin (red lines). For easy of comparison, we also plot the same mean detection efficiencies for all stars in the whole sample (black lines). 
\label{figDETDD}}
\end{figure*}

\begin{figure*}[!h]
\centering
\includegraphics[width=0.6\textwidth]{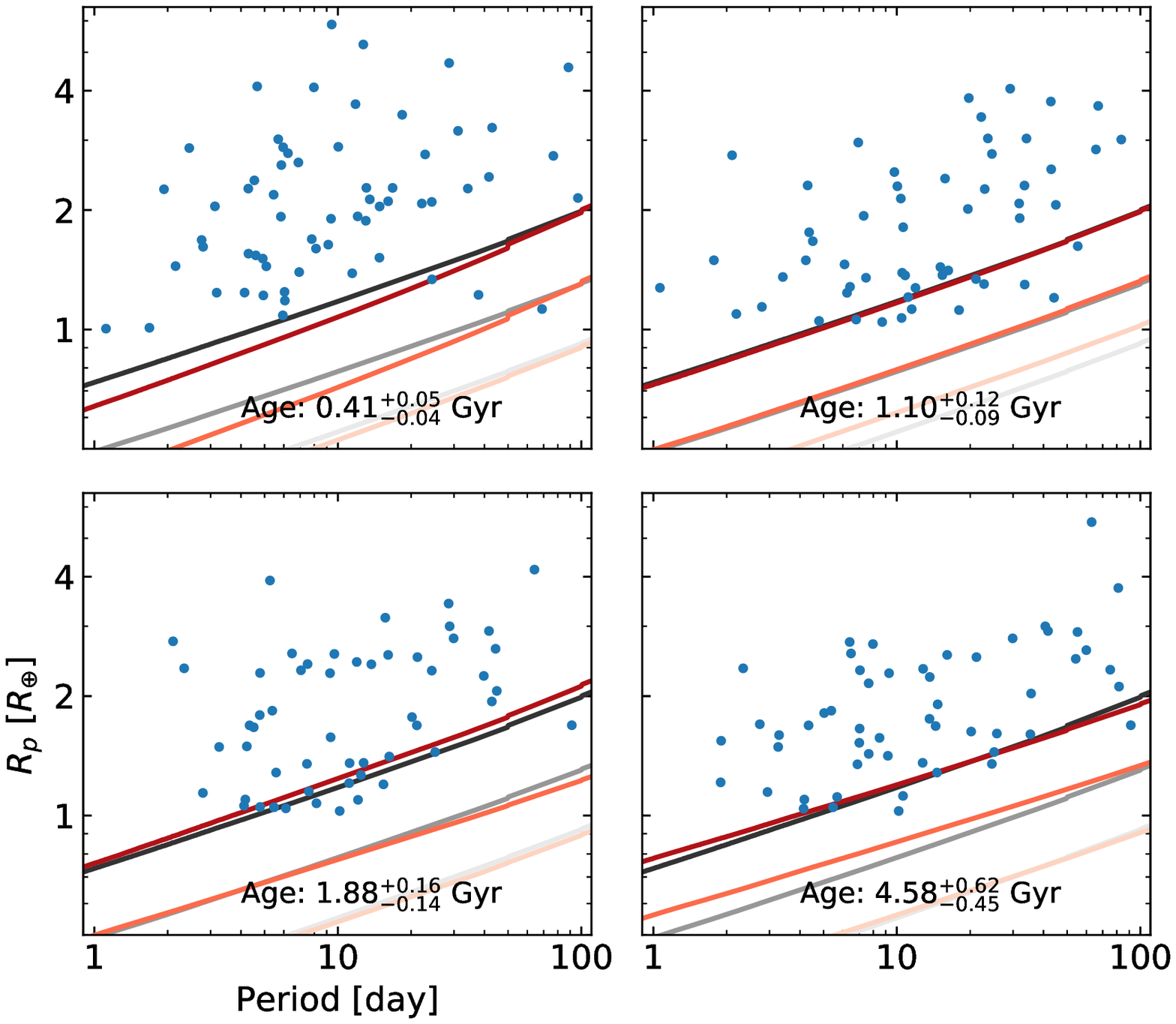}
\caption{Similar to Figure \ref{figDETDD} but the four bins of different age. 
\label{figDEAge}}
\end{figure*}

\begin{figure*}[!h]
\centering
\includegraphics[width=0.6\textwidth]{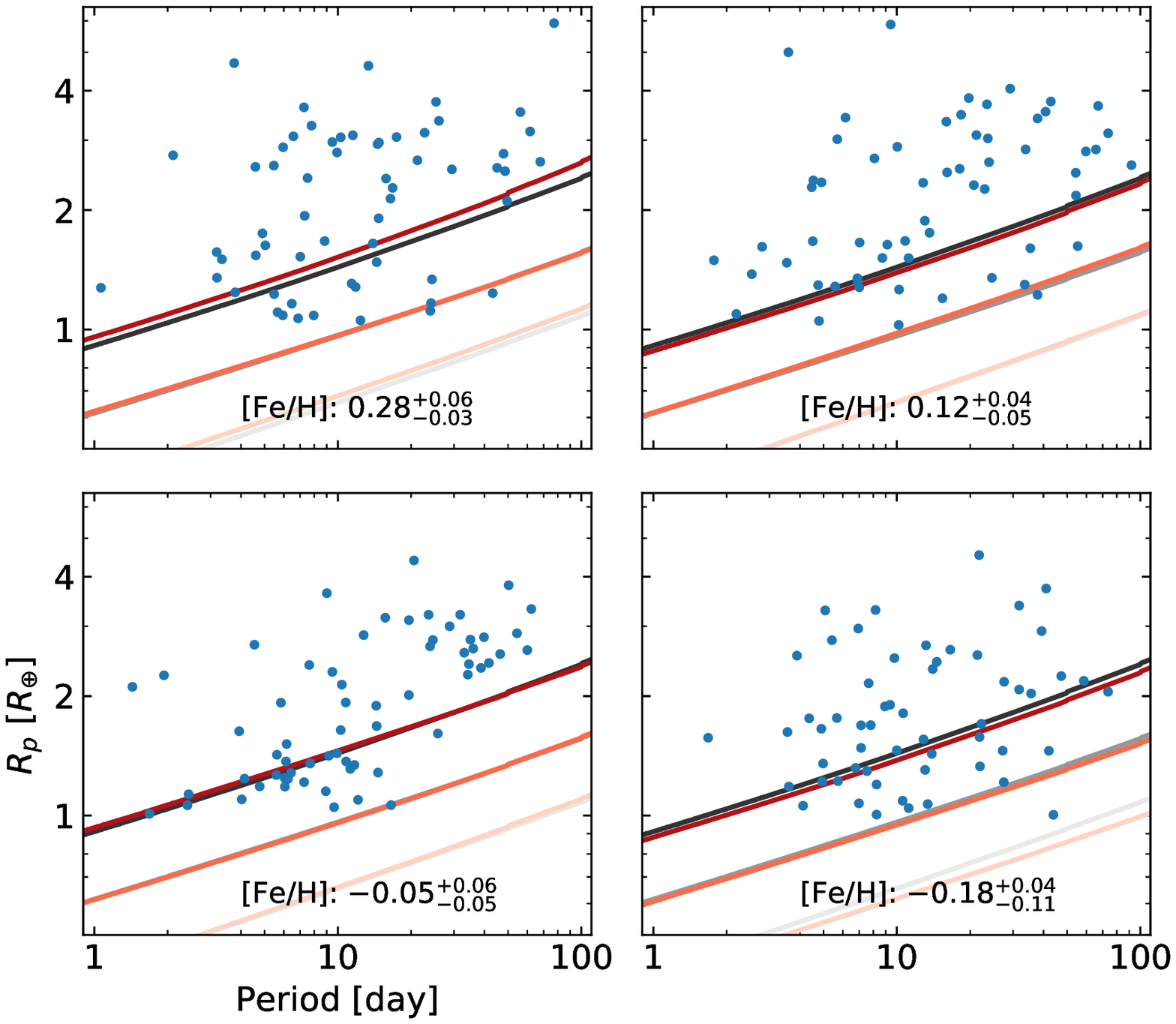}
\caption{Similar to Figure \ref{figDETDD} but the four bins of different $\rm [Fe/H]$. 
\label{figDEFeH}}
\end{figure*}

\begin{figure*}[!h]
\centering
\includegraphics[width=0.6\textwidth]{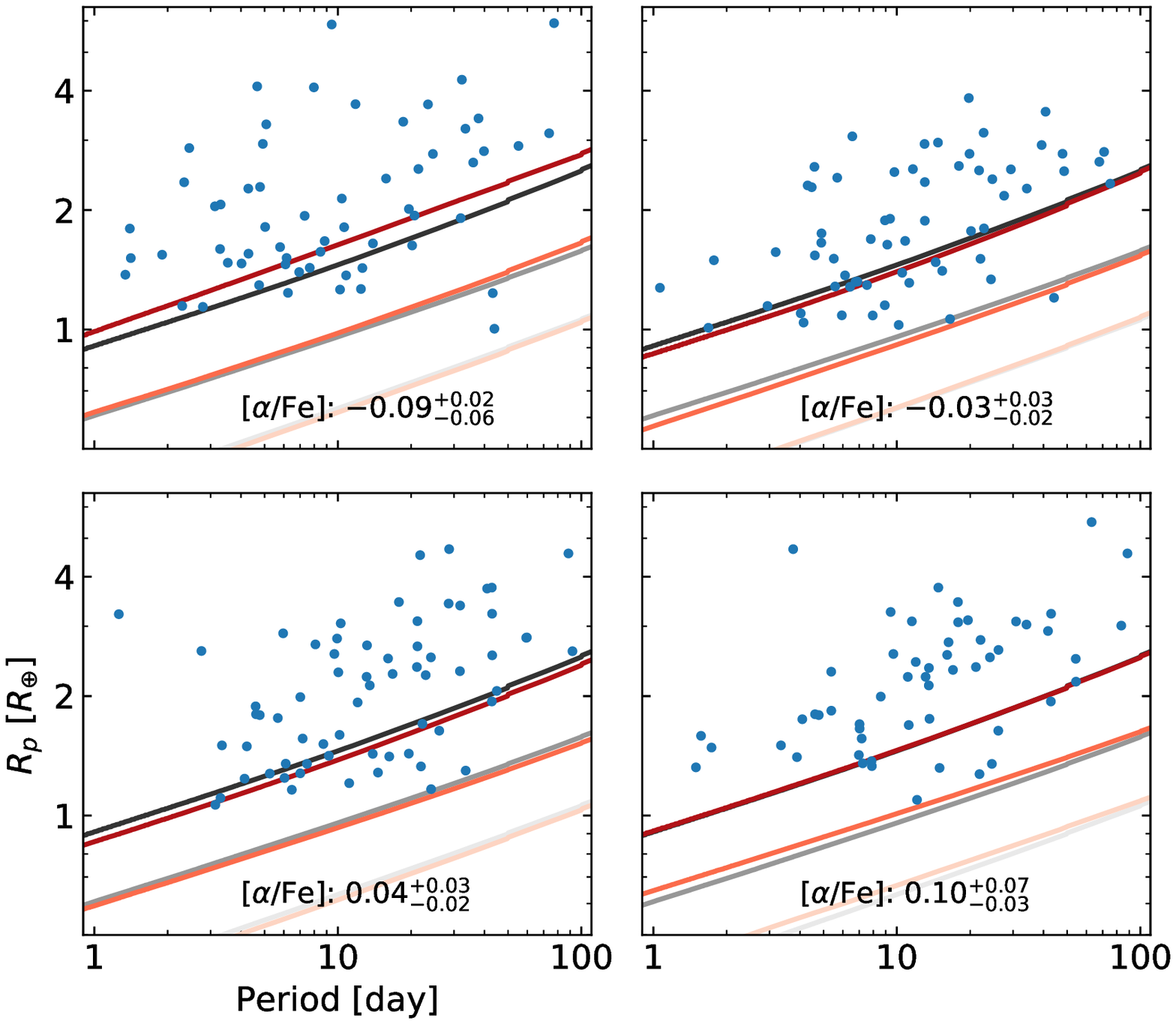}
\caption{Similar to Figure \ref{figDETDD} but the four bins of different $\rm [\alpha/Fe]$. 
\label{figDEAlpha}}
\end{figure*}

{\cdc We calculate the transit detection efficiency using the KeplerPORTs \citep{2017ksci.rept...17B} and the detection metrics available from the NASA exoplanet archive (https:// exoplanetarchive.ipac.caltech.edu/docs/). 
Figure \ref{figDETDD}, \ref{figDEAge}, \ref{figDEFeH} and \ref{figDEAlpha} show the 10\%, 50\%, and 90\% average detection efficiency contours as well as the tranet distributions in the period–radius diagram for the four bins of different $TD/D$, kinematic age, $\rm [Fe/H]$ and $\rm [\alpha/Fe]$, respectively.

As can be seen, for all the cases, most of planets after parameter controlling lie above the 90\% completeness curve and nearly all planets are above the 50\% completeness curve. Furthermore, for the bins of a given studied property (i.e., $TD/D$, age, $\rm [Fe/H]$ and $\rm [\alpha/Fe]$ , the detection efficiencies are close to each other.
Based on the above analysis, we conclude that our results will not be (significant) changed after completeness correction.}

\clearpage
\setcounter{figure}{0}
\section{Correlations between planetary and stellar properties for the CKS \uppercase\expandafter{\romannumeral7} sample with and without parameter control}

In the section \ref{sec.res.FeH} of main context, using the LAMOST-Gaia-Kepler catalog, we find that the average radii of planets larger than valley planets ($2.1-6 R_\oplus$) increase with stellar metallicity after controlling other properties (e.g., stellar mass, age and etc.). 
However, with the CKS \uppercase\expandafter{\romannumeral7} sample \citep{2018AJ....156..264F}, recently \cite{2021AJ....162...69K}
showed that the observed correlation between stellar metallicity and planetary radius could be explained by the combination of the stellar mass-metallicity correlation and the stellar mass-planetary radius correlation, which is inconsistent with our results.

\begin{figure*}[!h]
\centering
\includegraphics[width=0.75\textwidth]{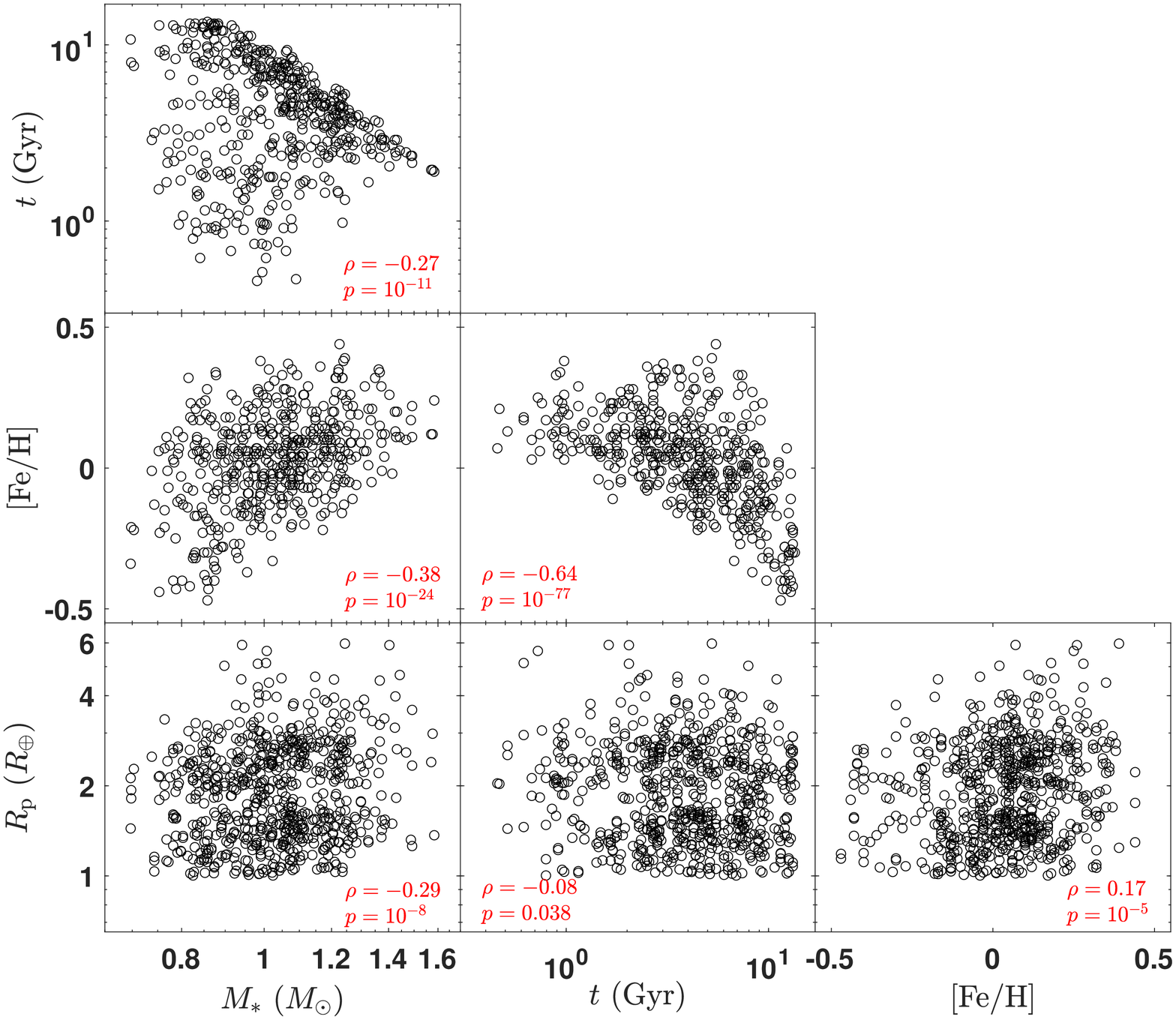}
\caption{Correlations between planetary and stellar properties for the CKS \uppercase\expandafter{\romannumeral7} sample \emph{before} parameter control.
In each panel, $\rho$ and $p$ denote the Pearson correlation coefficient and  $p-$value for each pair of variables.
\label{figCKS7SPC}}
\end{figure*}

To understand the reasons for this discrepancy, we conducted similar correlation analyses with CKS \uppercase\expandafter{\romannumeral7} sample as in \cite{2021AJ....162...69K}. 
As can be seen in Figure \ref{figCKS7SPC}, we reproduce the correlations of $M_*-\rm  [Fe/H]$, $M_*-\rm [Fe/H]$ and $M_*-R_{\rm p}$ as seen in Figure 1 of \cite{2021AJ....162...69K} quantitatively.
Nevertheless, as shown in Figure \ref{figCKS7SPC}, we also find more correlations which were not considered in \cite{2021AJ....162...69K}: the stellar age is anti-correlated with both stellar mass and metallicity.
Therefore, these \emph{apparent} correlations are interrelated and should be taken with caution. 

\begin{figure*}[!h]
\centering
\includegraphics[width=0.65\textwidth]{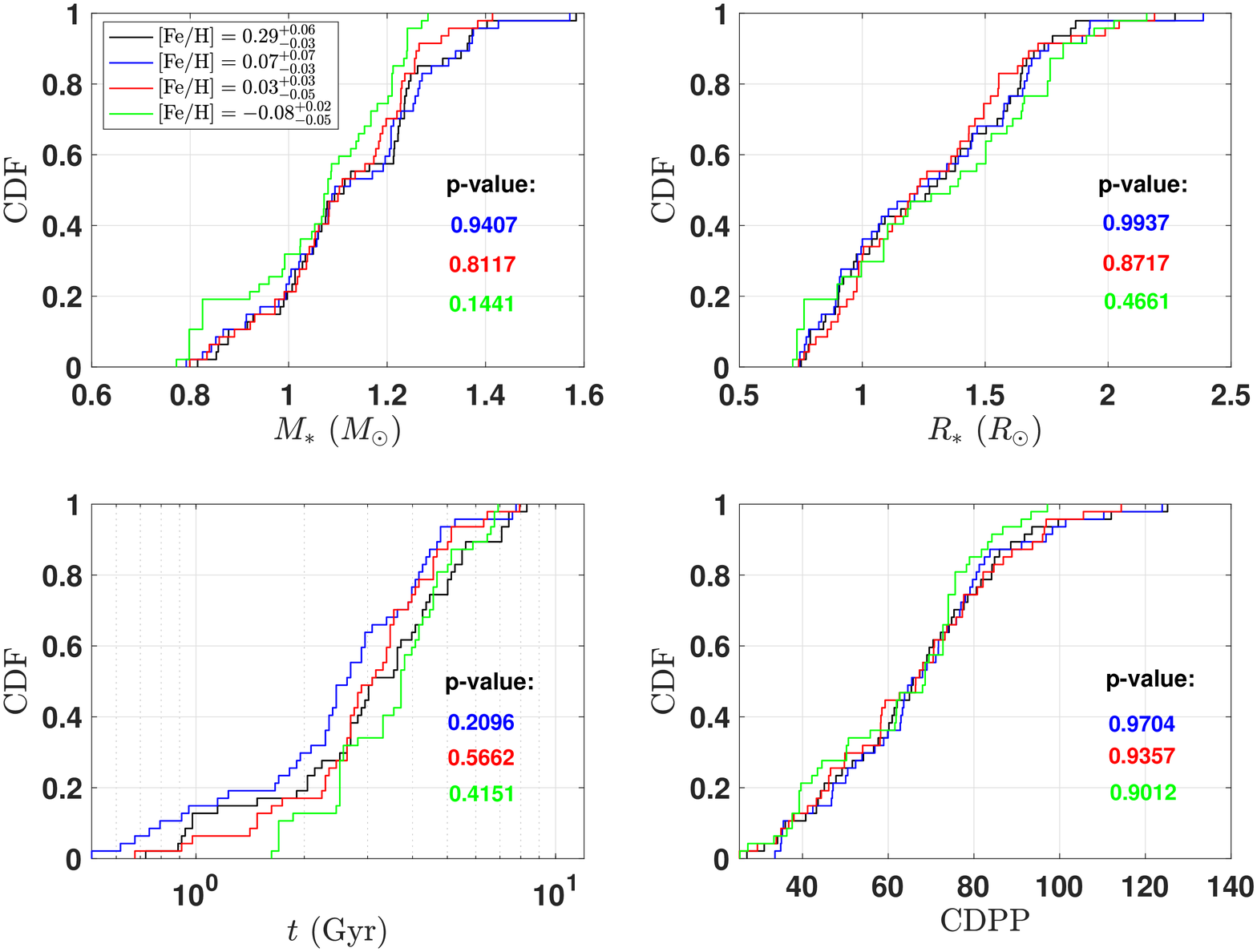}
\caption{The cumulative distributions of stellar mass $M_*$, radius $R_*$, age and CDPP 4.5h for the four bins with different metallicity $\rm [Fe/H]$ of CKS \uppercase\expandafter{\romannumeral7} \emph{after} parameter control. 
In each panel, the $p$ denotes the $p-$value of the two sample KS test for the distributions of neighboring star belonging to the latter three bins comparing to the first bins. 
\label{figCDFMRCDPPFeHCKS7}}
\end{figure*}

\begin{figure*}[!t]
\centering
\includegraphics[width=0.7\textwidth]{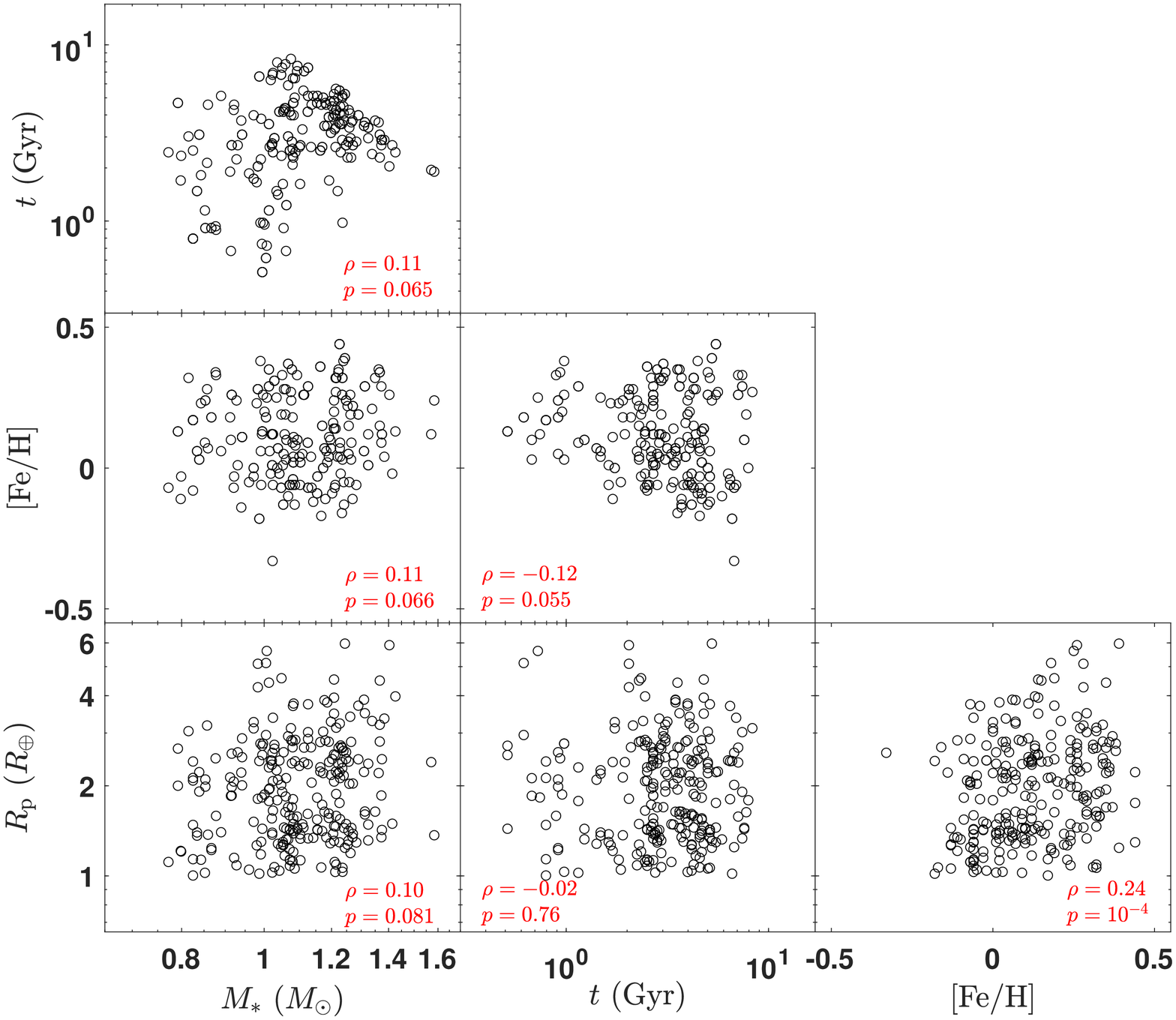}
\caption{Similar to Figure \ref{figCKS7SPC} but \emph{after} controling parameters (i.e., stellar mass, radius, age and CDPP) to let them have similar distributions as shown in Figure \ref{figCDFMRCDPPFeHCKS7}.
\label{figCKS7SPCAPCFeH}}
\end{figure*}

To isolate the effect of metallicity, following the same parameter control procedure as in section \ref{sec.res.FeH}, we divided the whole sample into four bins according to their $\rm [Fe/H]$ and used the Nearest Neighbour method to select the nearest neighbor in the space of the controlled parameters (i.e., stellar mass, radius, age and CDPP) from stars in the latter three bins for every star in the first bin. 
As shown in Figure \ref{figCDFMRCDPPFeHCKS7},  the distributions in mass, radius and age after parameter control are similar for stars in different $\rm [Fe/H]$ bins.
Then we compute the Pearson correlation coefficients and $p$ values for 
the mutual correlations among stellar mass, radius, age and planetary radii.
As shown in Fig. \ref{figCKS7SPCAPCFeH}, the correlations among stellar parameters become statistically insignificant with $p$ values larger than 0.05. 
In contrast, the correlation between $\rm [Fe/H]$ and $\rm R_p$ still maintains statistically significant with a $p$ value $=10^{-4}$, which is consistent with our results using the LAMOST-Gaia-Kepler sample.

Above exercises demonstrate that it is crucial to extract the \emph{intrinsic} correlations (e.g., via parameter control) rather than using the \emph{apparent} correlations among a number of planetary and stellar properties.

\clearpage

\bibliography{library.bib}

\end{document}